# High Performance Electrostatic Sensors and Actuators for LISA Proof Mass Control.

*Giorgio Fontana*

*University of Trento*

*dep. Of Materials Engineering*

# Abstract.

This document contains two presentations which describe the working principles of a class of electrostatic multidimensional sensors and force actuators.
The subject of the study is the search of the most effective methods for measuring the position of a cubical conducting proof mass which floats in a weightless environment.
The same proof mass must be controlled with a feedback loop by applying forces with the same set of electrodes.
For more information please see the web site:

http://lisa.jpl.nasa.gov/

# Principles of multidimensional contactless capacitive AC sensors.

*Giorgio Fontana*

*University of Trento*

*dep. Of Materials Engineering*

*June. 2001*

# Sensors.

## Bridge for one degree of freedom.

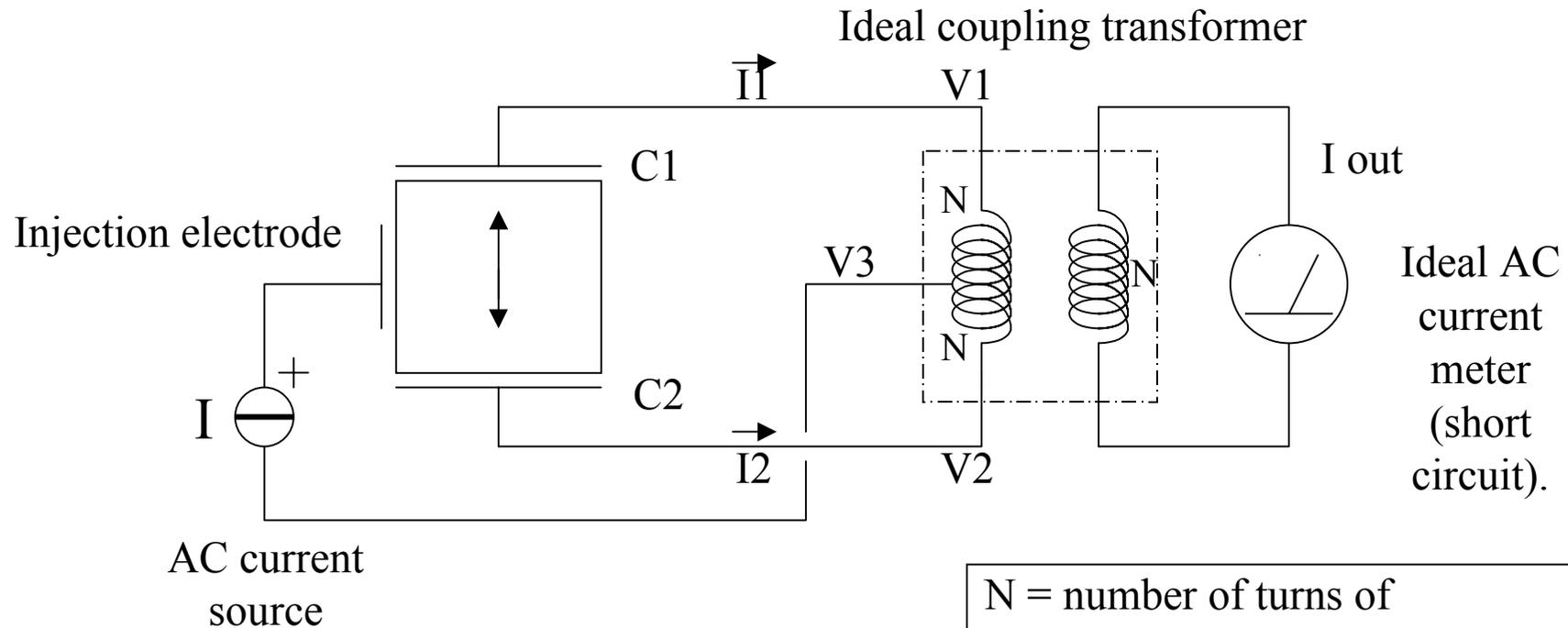

An injection electrode is required.
*The injection electrode is an unwanted source of force and stiffness.*

$N$ = number of turns of the ideal transformer

$V1 = V2 = V3$ (output winding shorted)

$I_{out} = I(C1-C2)/(C1+C2)$, with p.s.d.

# The bidimensional bridge.

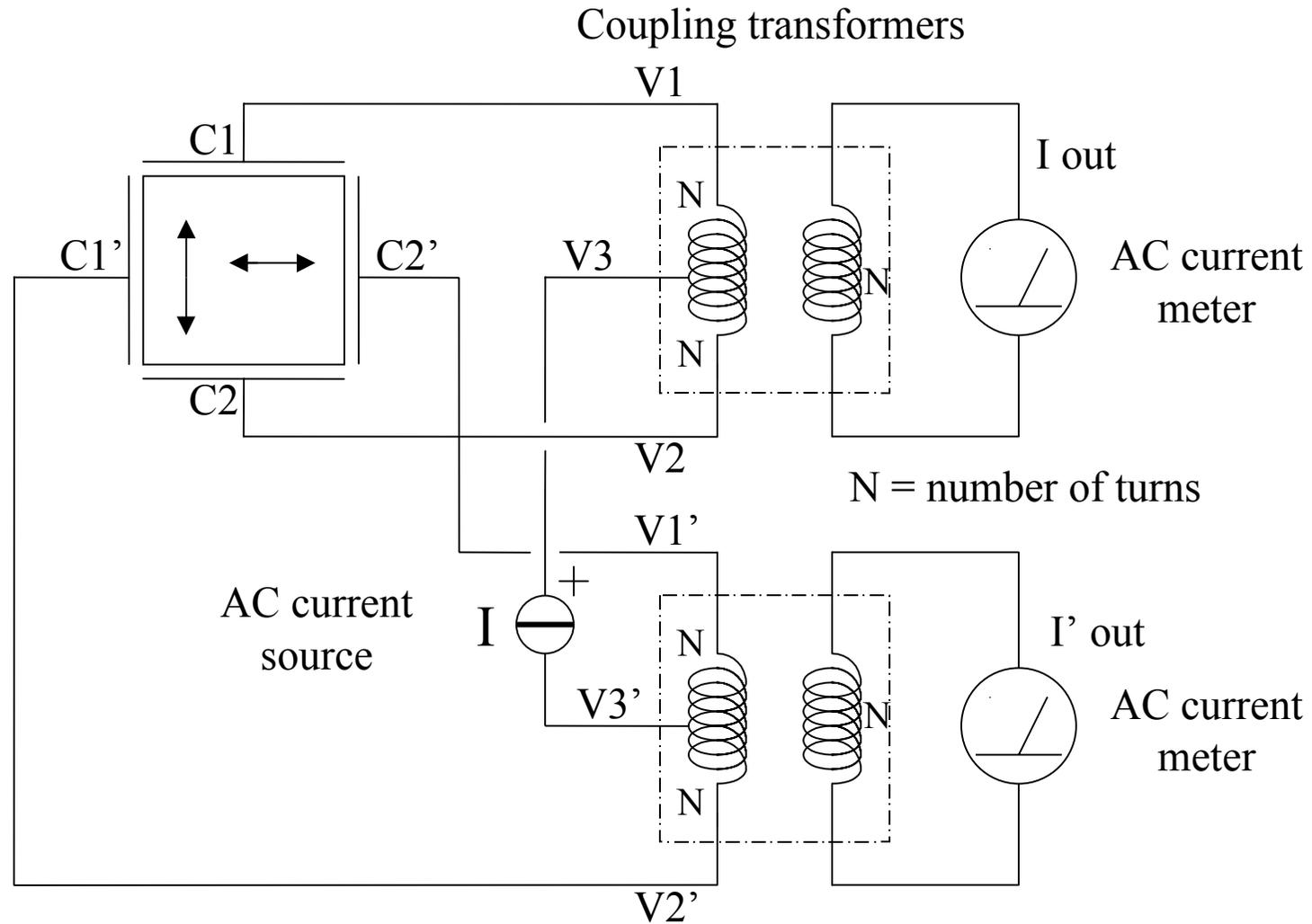

An injection electrode is not required. THIS IS THE MAIN GOAL.

*Equations: the same as preceding slide, with and without '.*

# The bidimensional bridge, computing the voltage drop.

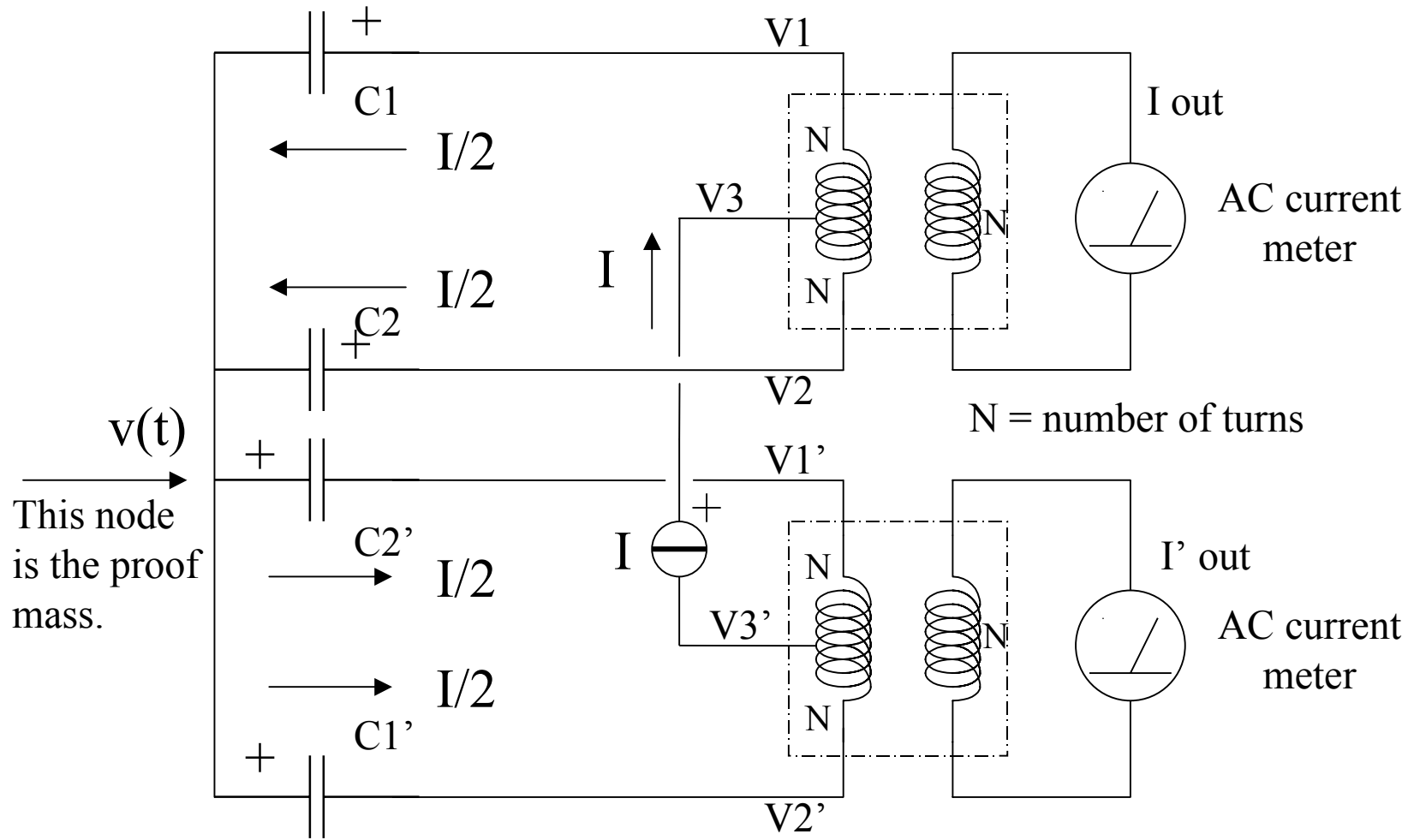

By definition of current measurement V3=V1=V2 and V3'=V1'=V2'.
With C1=C2=C1'=C2'=C and $v_C(0-)=0$ (by UV discharge) we have
$V(\omega)$ across the generator I => (V3−V3') = $I/(\omega C)$
$V(\omega)$ across C => $V_C = I/(2\omega C)$, $I_C = I/2$ and (v3(t)+v3'(t))/2=v(t)

## The bidimensional bridge, definition of the potentials.

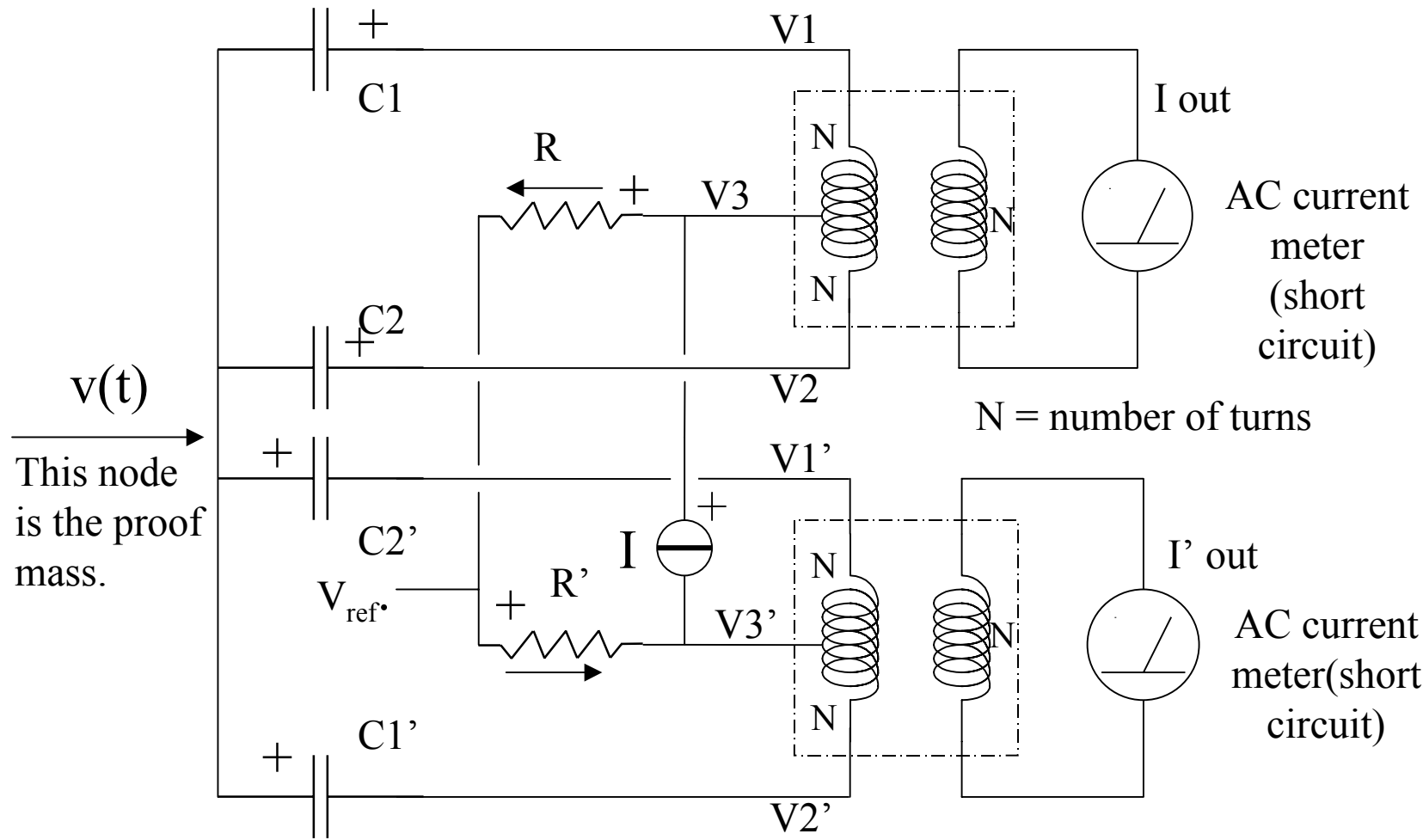

Potentials **MUST** be defined with an additional network (R=R').
giving $(v3(t)+v3'(t))/2 = v_{ref}(t)$, because $I_R = I_{R'}$ and $V_R = V_{R'}$.

# The bidimensional bridge, potential of the cube as a function of time.

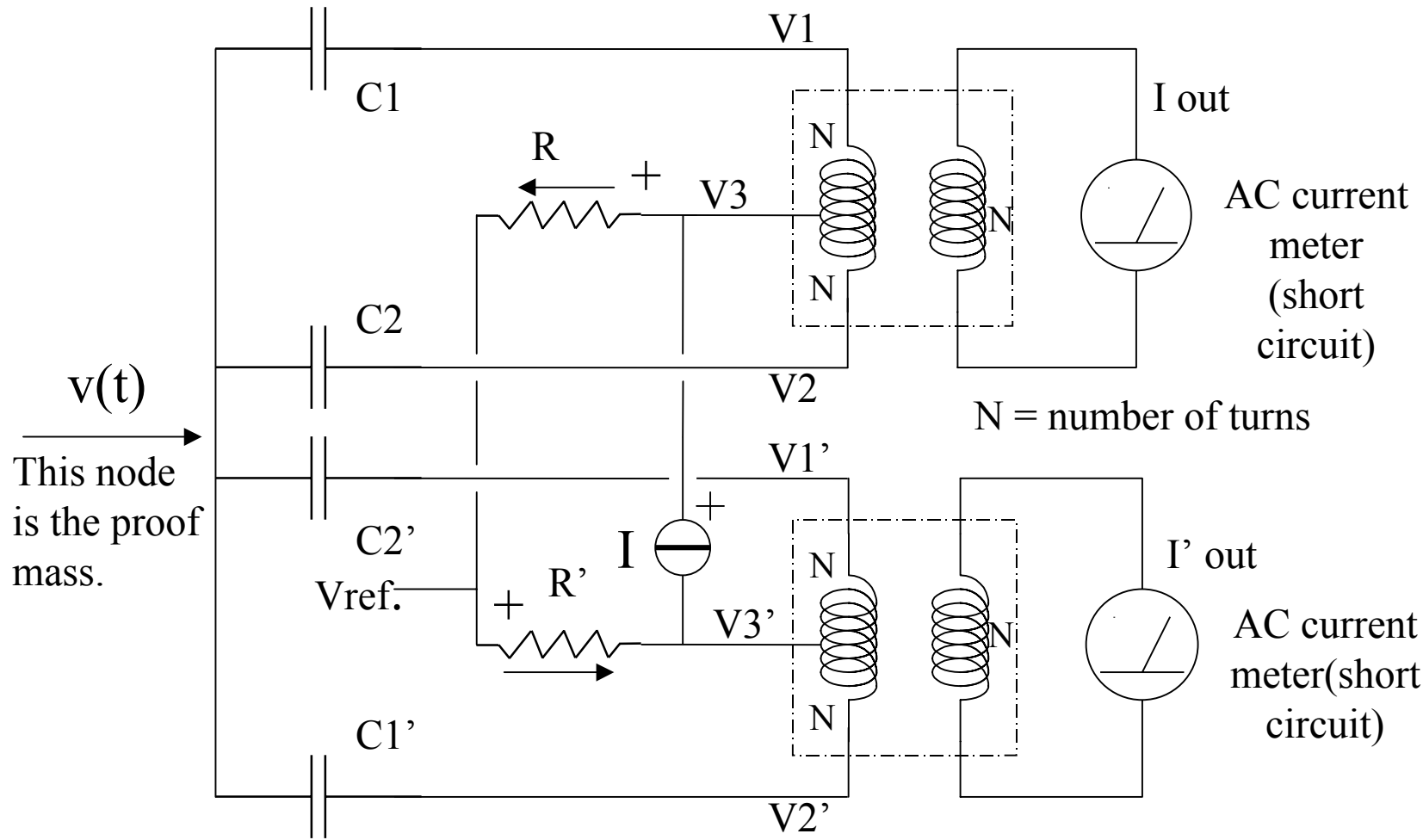

By comparison we have v(t) =$v_{ref}$(t), and with $v_{ref}$(t)≡0
we have v(t) ≡0.    HERE COMES THE EXTRA BONUS!
**The proof mass is at the reference potential (0) $\forall$ t.**

Not convinced ? The equivalent circuit for the potentials:

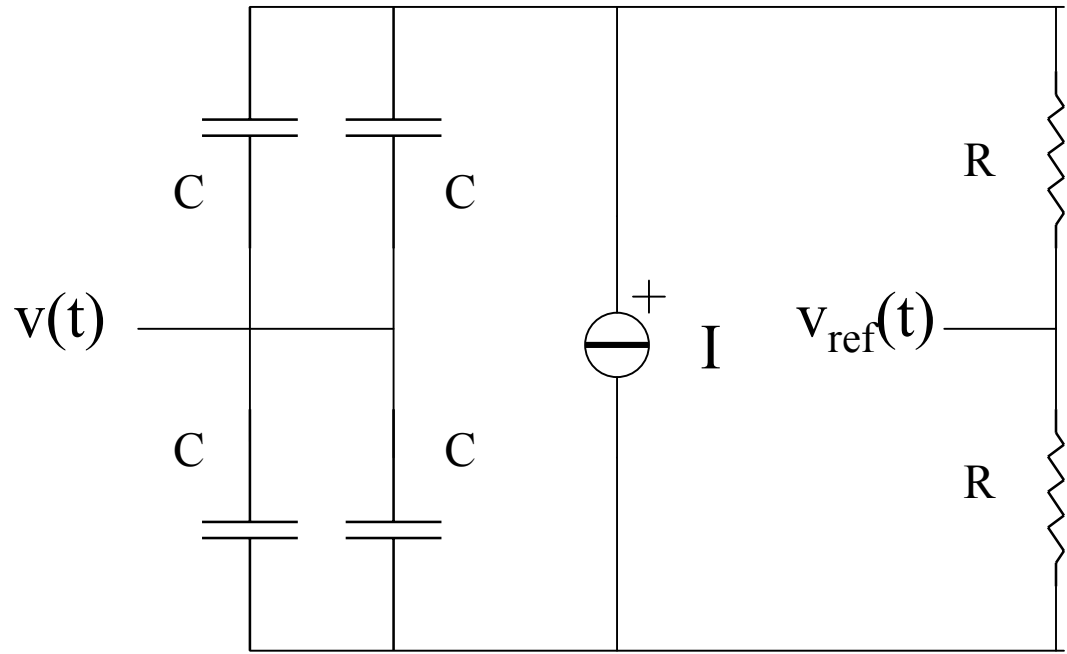

At the equilibrium, i.e. with the values shown in the above schematic, $v(t) = v_{ref}(t)$ = potential of the proof mass, because the bridge is balanced.

For DC balance R must be as high as possible, but much less than stray resistances.
For AC balance R must have a capacitor in parallel to compensate for stray capacitances in transformers.

## What happens with a "real" current source?

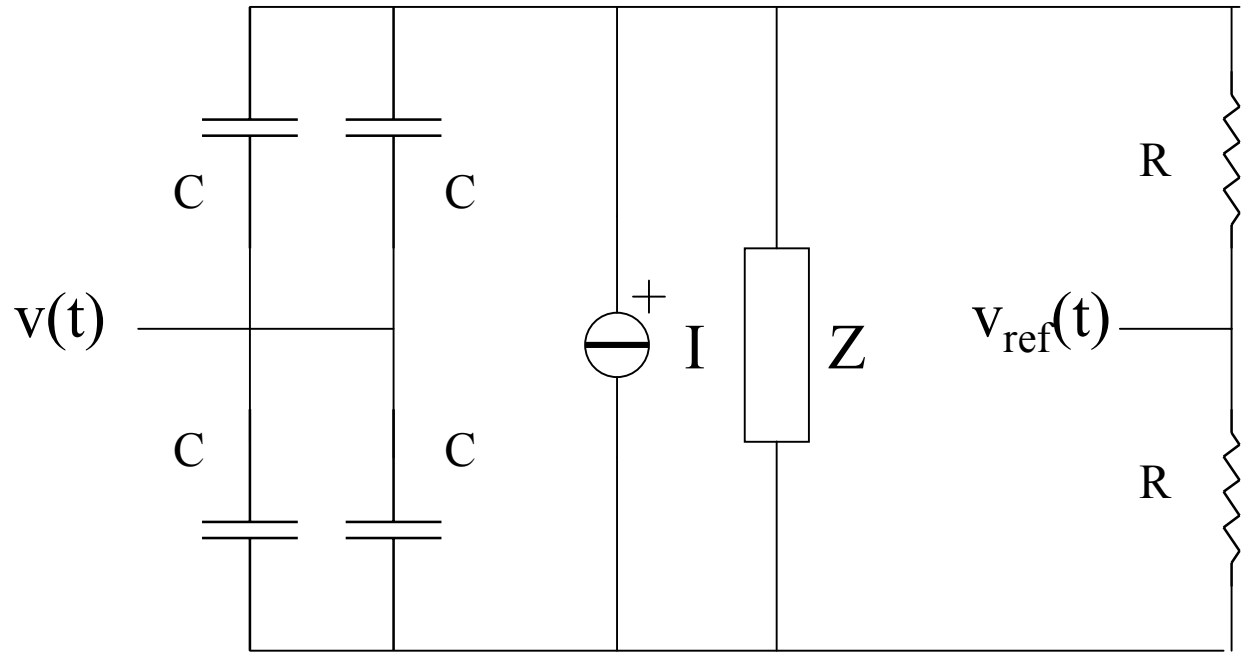

NO EFFECT ON v(t).

## Switching to a "real" equivalent voltage source.

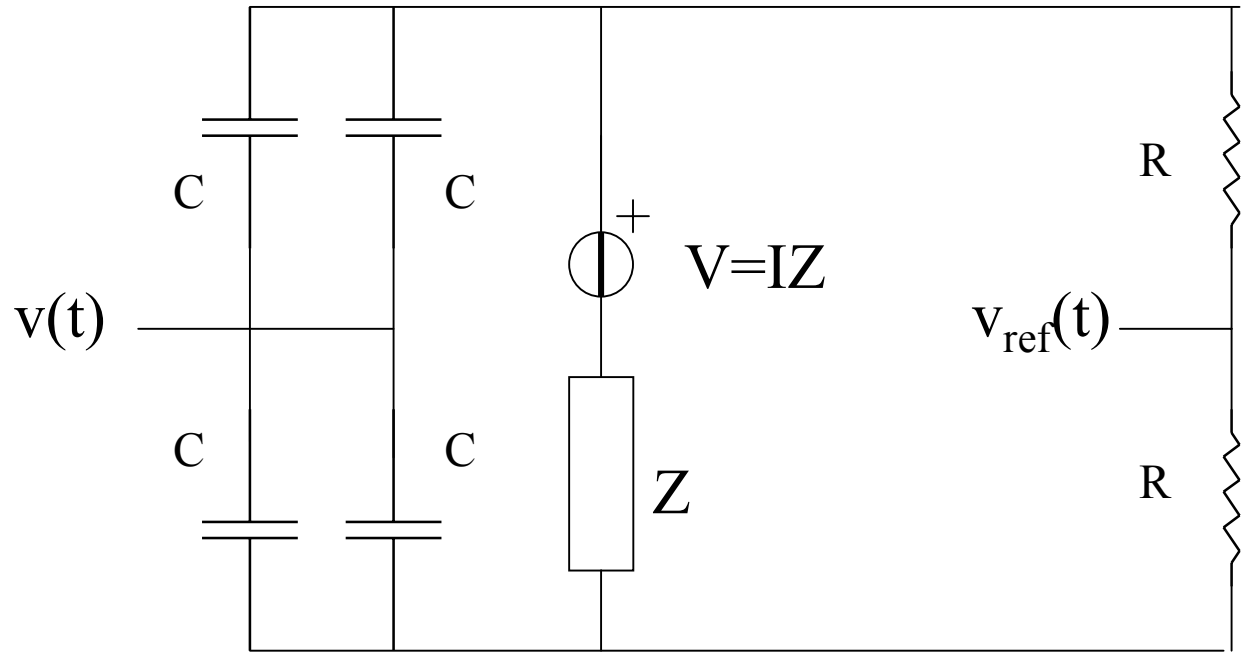

NO EFFECT ON v(t).

## Adding "real" bridge transformers.

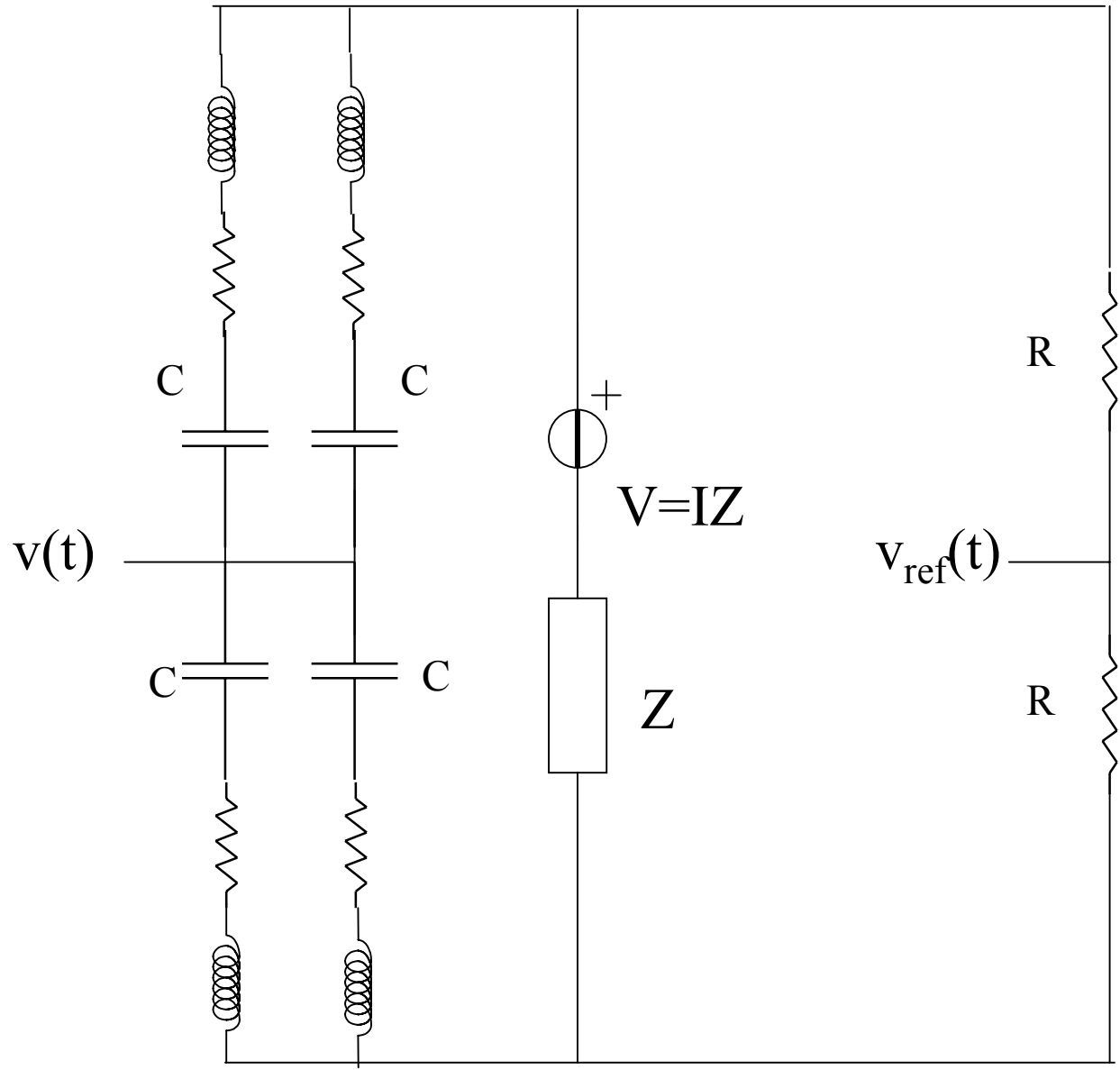

NO EFFECT ON v(t) (with equal transformers).

# 2Ndimensional bridges principles.

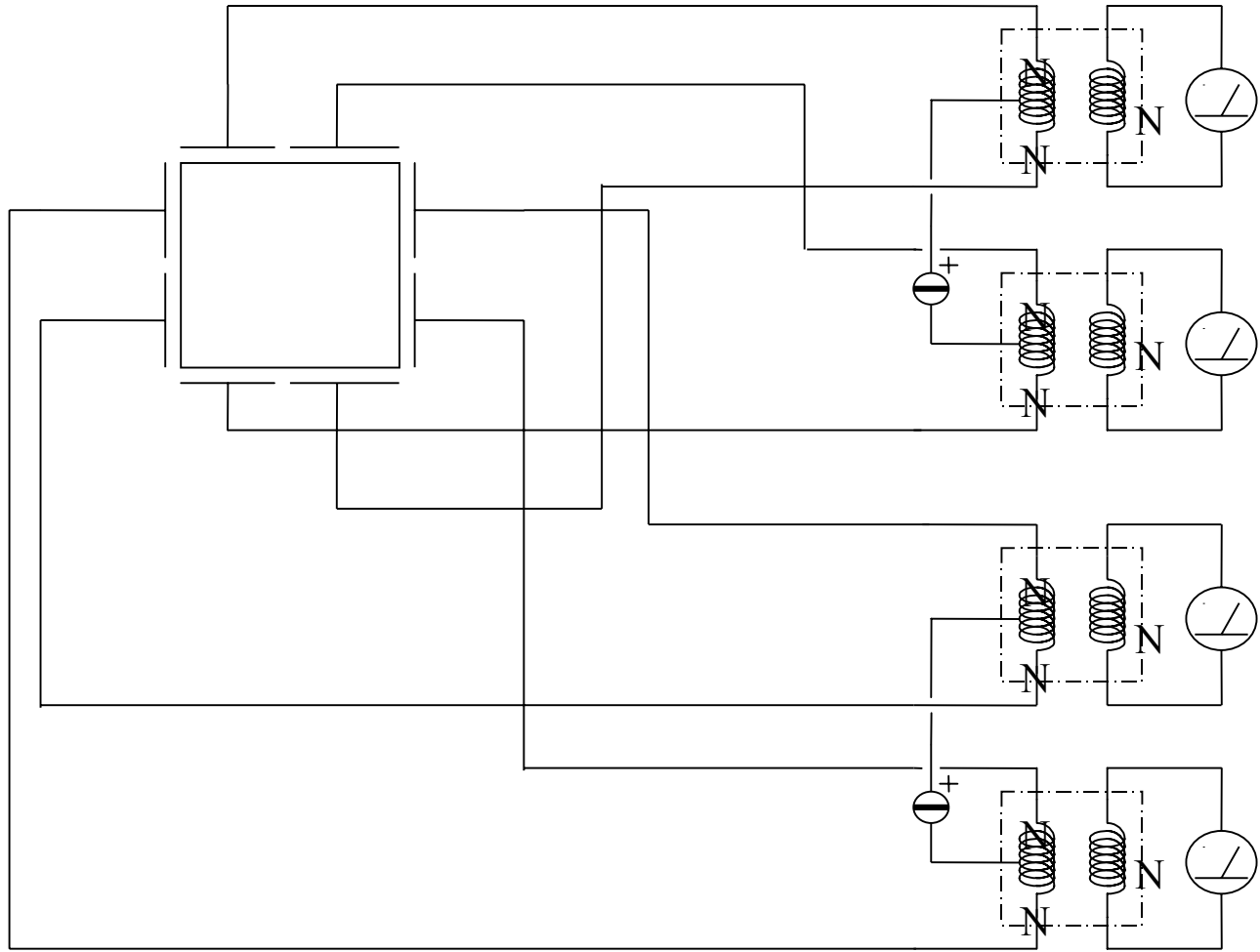

**Equations**: *the same as the bidimensional bridge. Current sources may have the same frequency because of charge conservation. All $v_{ref}$ must be the same for zero applied force.*

# Circuit Morphing for Noise calculation.

With a real sine current generator.

# The Original configuration.

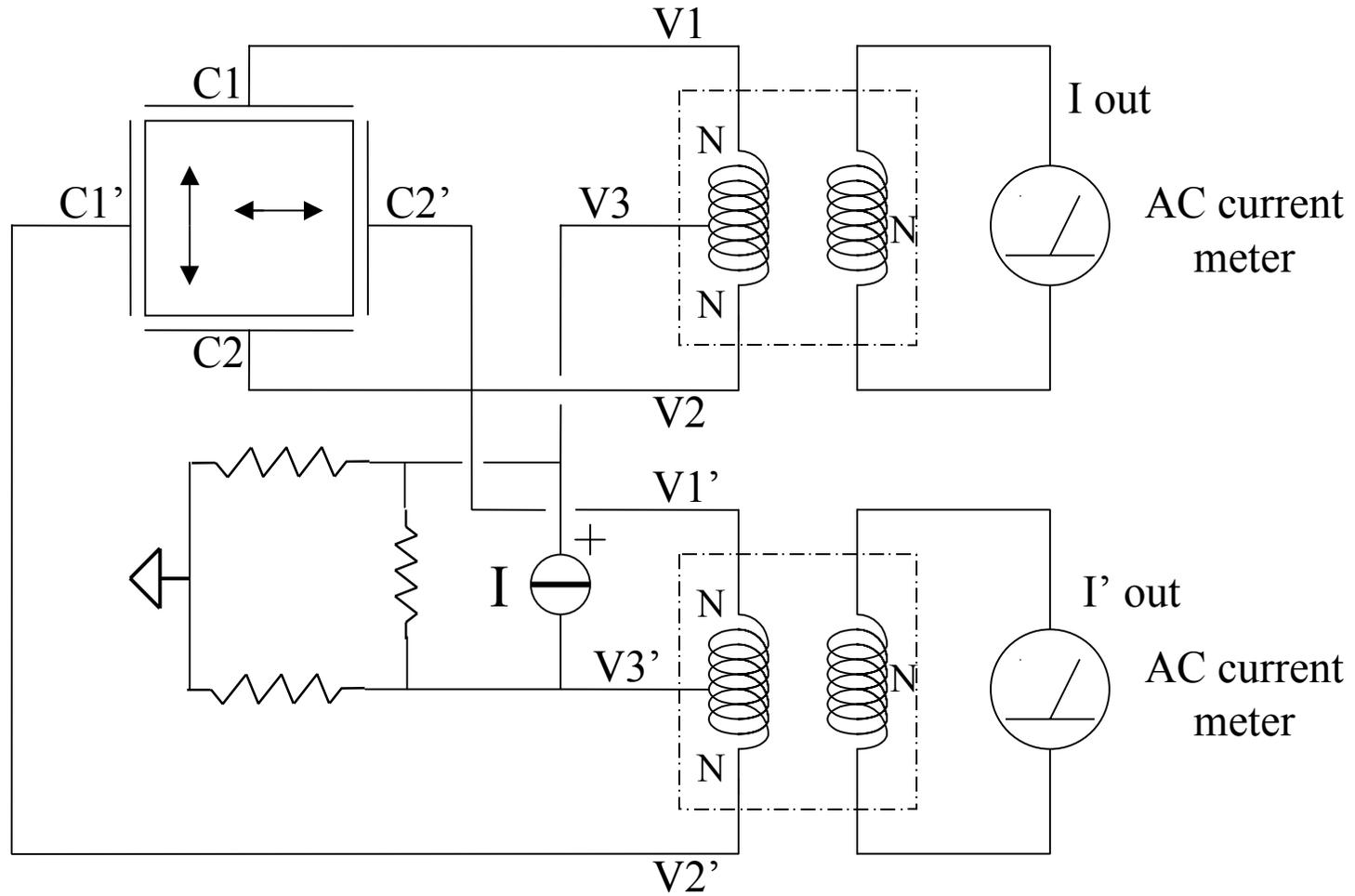

# STEP 1 - Removing a bridge.

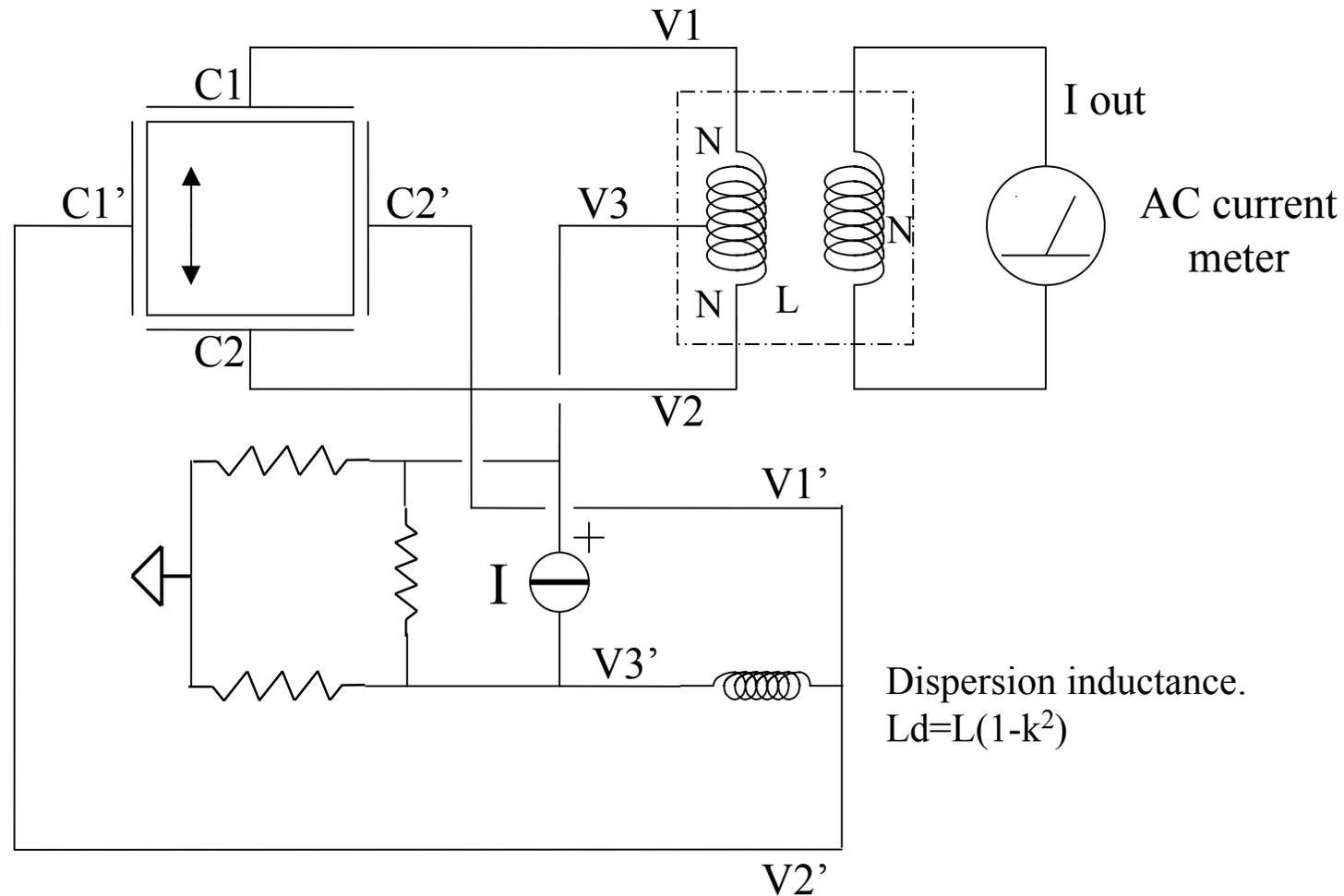

Consequences: no consequence on bridge noise behaviour, unfortunately an axis is no longer observable.

# STEP 2 - Thevenin equivalent circuit for the generator.

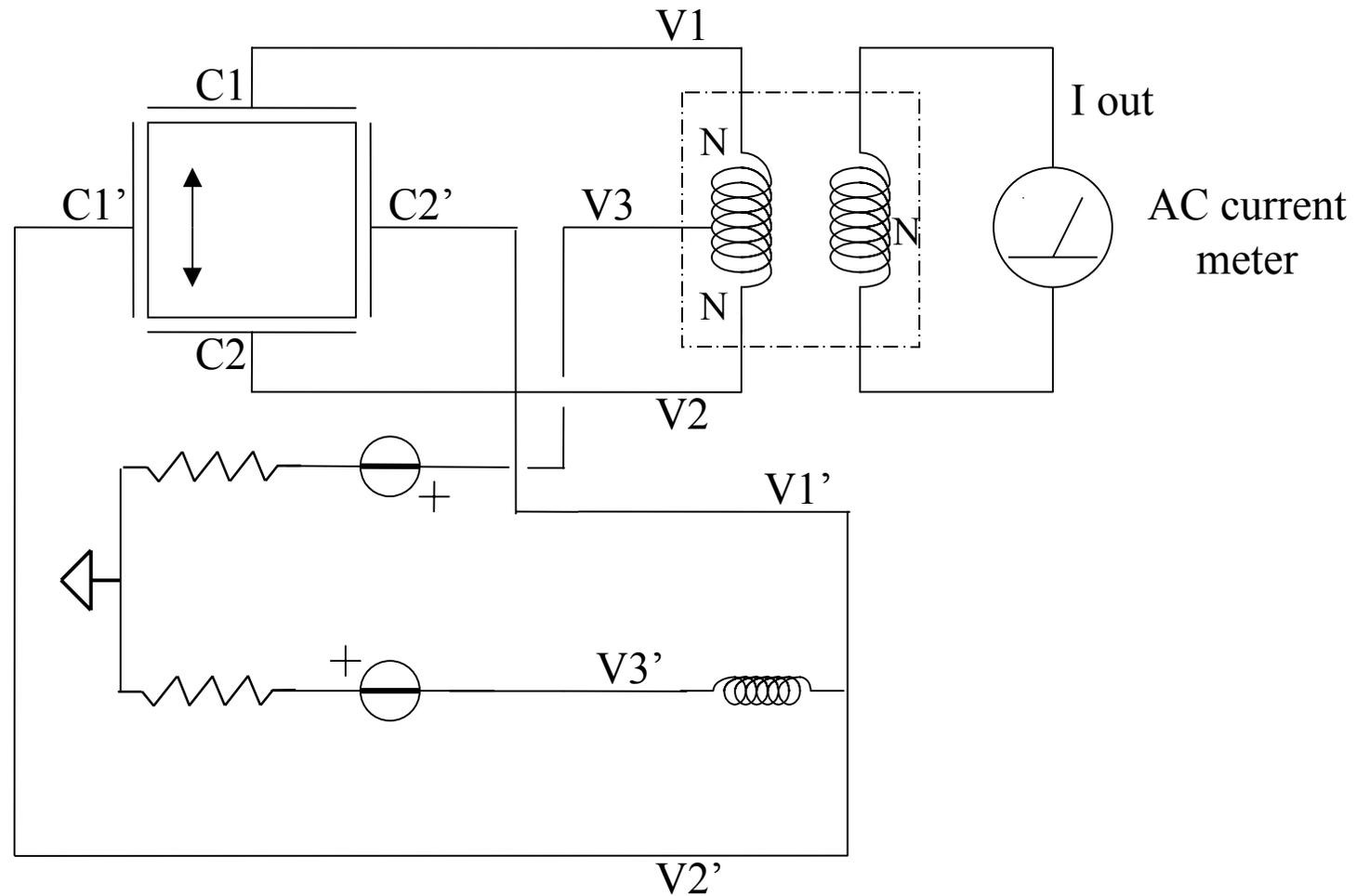

Consequences: no consequence on bridge noise behaviour.

# STEP 3 - Collapsing C1' and C2' to C'=C1'+C2'.

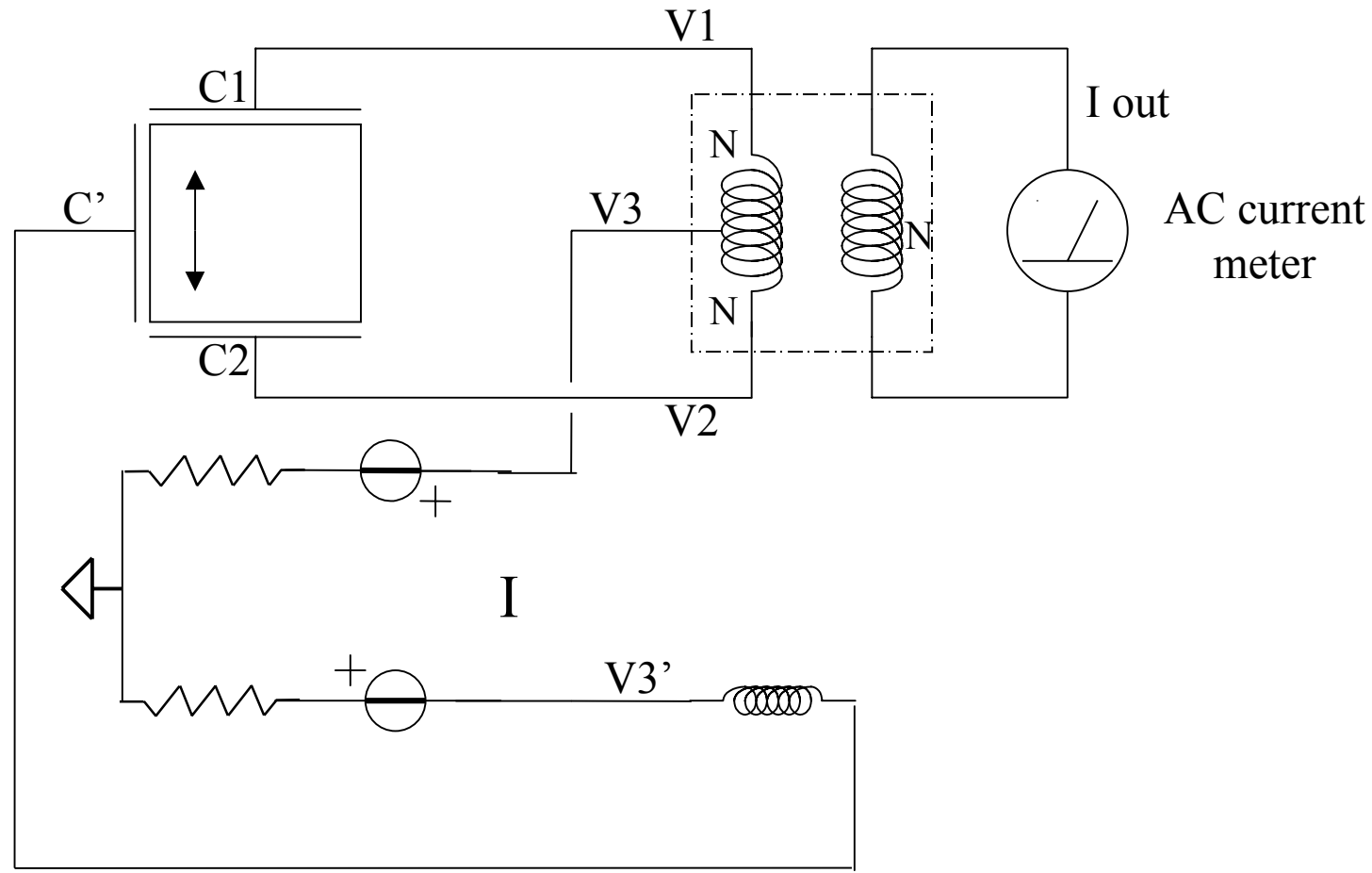

Consequences: no consequence on bridge noise behaviour.
X-axis back-action changed, possible higher induced amplitude
modulation of the generator current. The symmetry is lost.

# STEP 4 - Moving the "ground symbol".

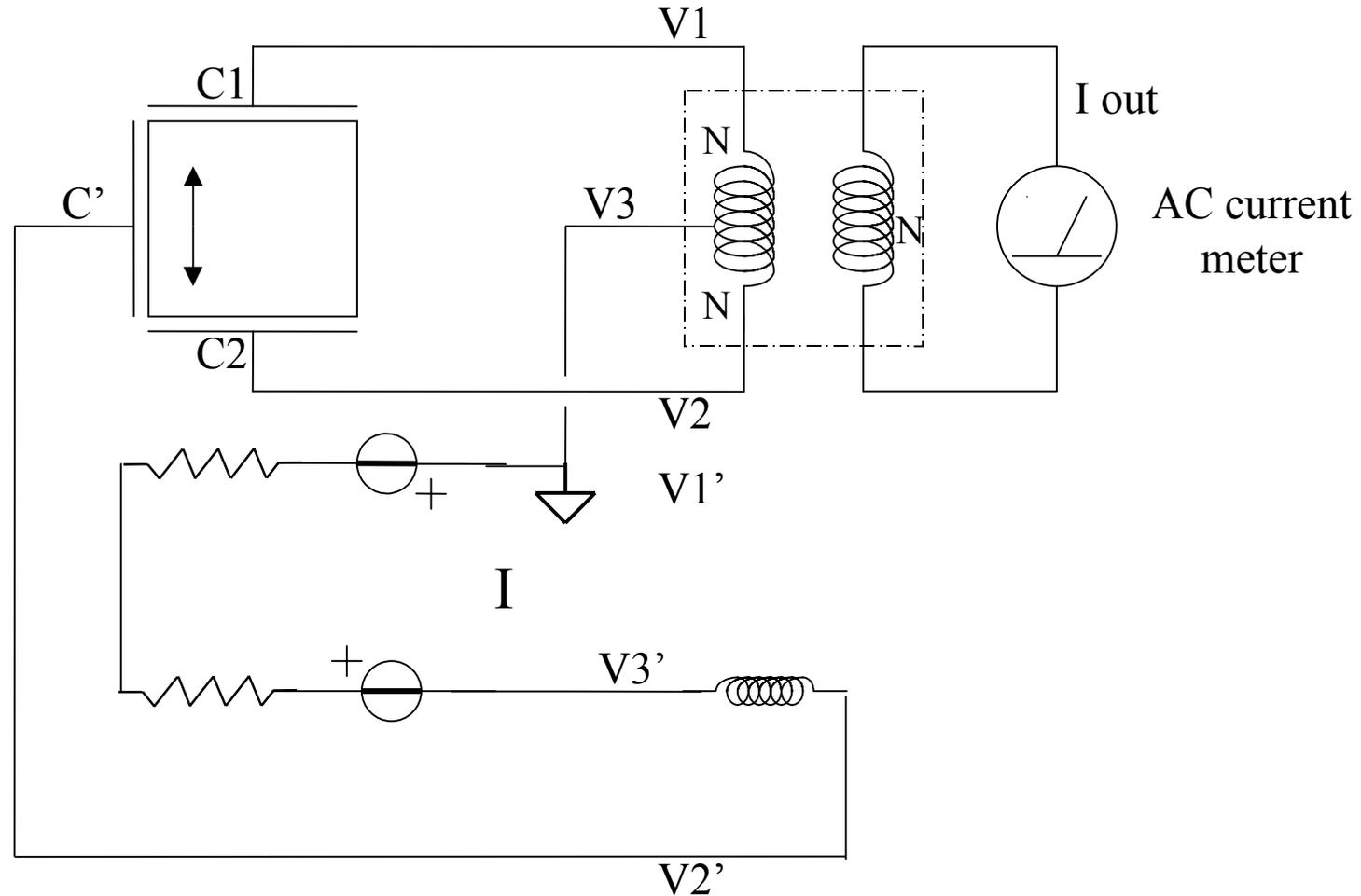

Consequences: no consequence on noise behaviour. The proof mass is no longer a virtual ground.

# STEP 5 - Collapsing of voltage sources

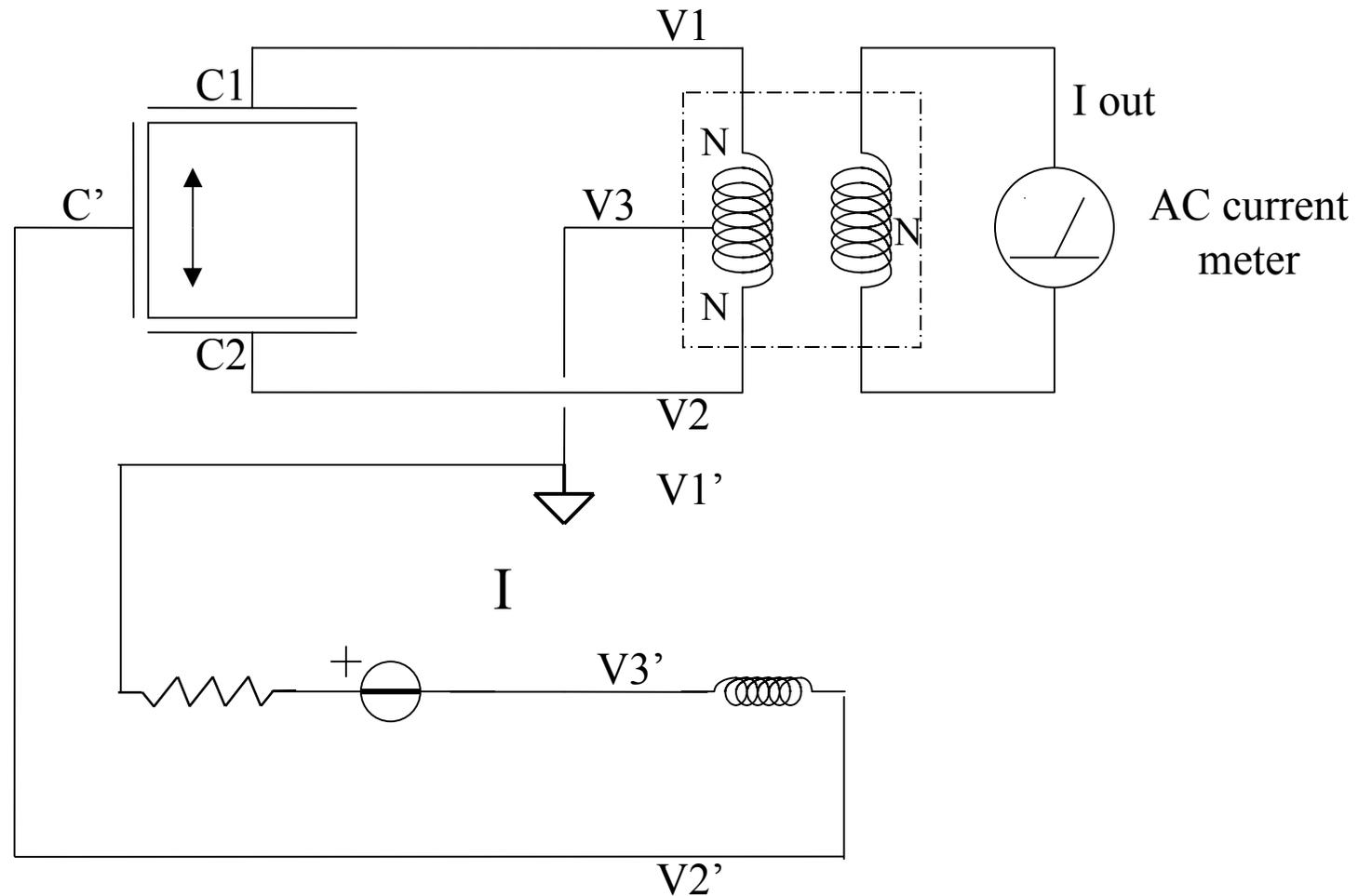

Consequences: no consequence on noise behaviour.

# STEP 6 - Some cosmetics.

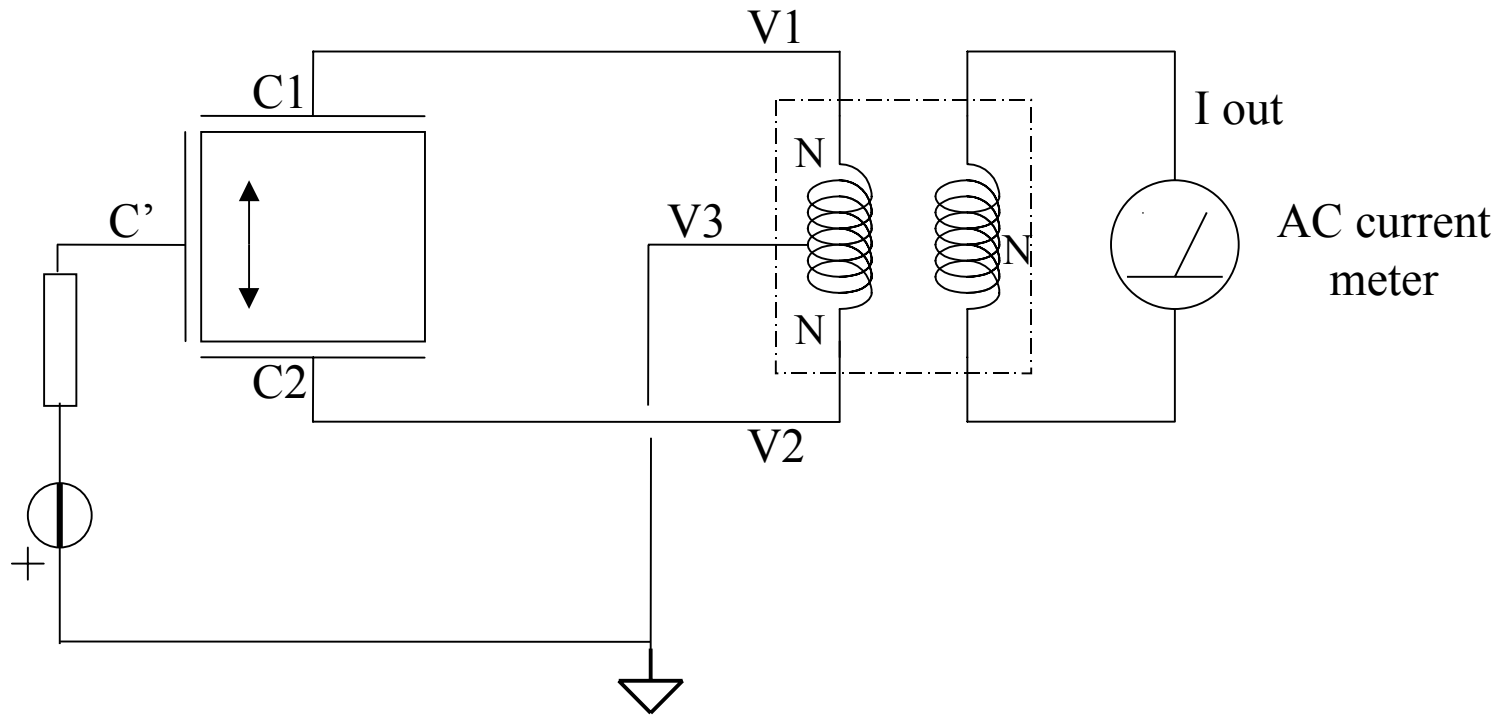

Consequences: no consequence on noise behaviour.

CONCLUSION.

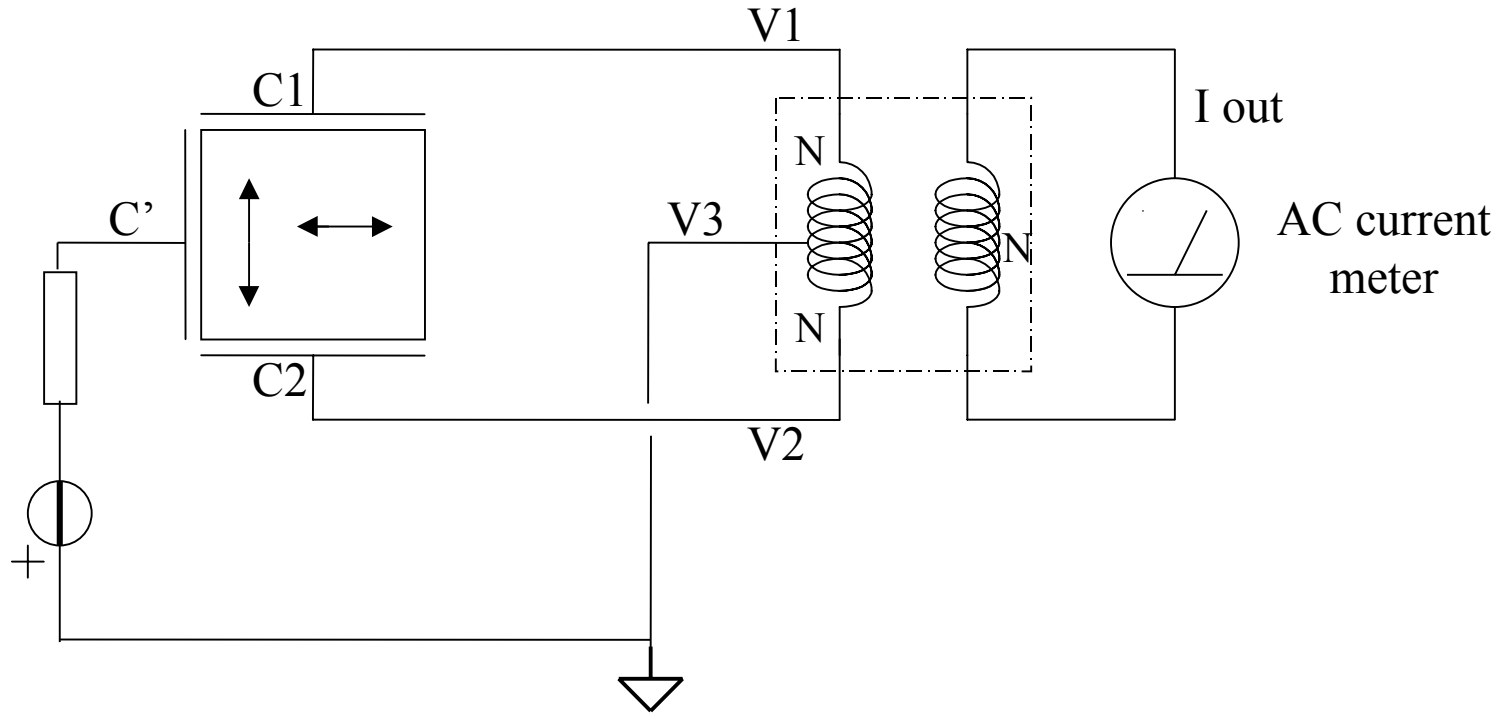

THE NOISE BEHAVIOUR OF THE DOUBLE BRIDGE CAN BE COMPUTED ANALYZING THE SINGLE BRIDGE.

OK, THE CALCULATION HAS BEEN ALREADY DONE!

# Advantages of the proposed configuration.

1) High symmetry.
2) No injection electrode.
3) No additional stiffness due to injection electrodes.
4) Readout electrode stiffness identical to conventional configuration.
5) Proof mass at zero (reference) potential ($\forall$ t) at the equilibrium.
6) Proof mass is a *virtual* ground.
7) Maximum use of proof mass area for measurement.
8) Equidistribution of forces due to measurement currents at the equilibrium.
9) Possible use of a single frequency for multidimensional sensors.

# Disadvantages of the proposed configuration.

1) High common mode on readout transformers (generator voltage).
2) Restriction on the number of bridges: only 2N bridges allowed.

# Counteracting the common mode.

1) Electrostatic shielding

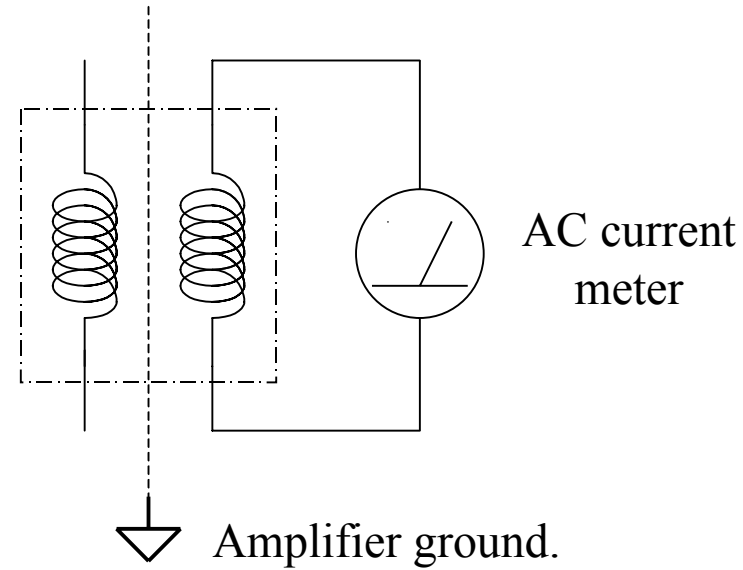

AC current meter

Amplifier ground.

2) Use of a BALUN (balanced/unbalanced)

Magnetic core.

The BALUN shown is a magnetic device with a high inductance for the common mode and a very low inductance for the differential mode. Values can be designed as needed.
The differential mode does not "see" the core.

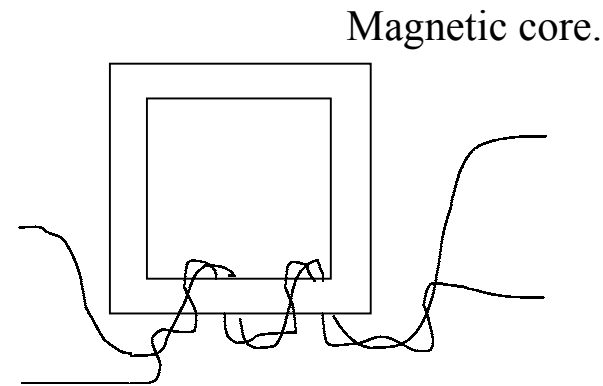

# A floating preamplifier.

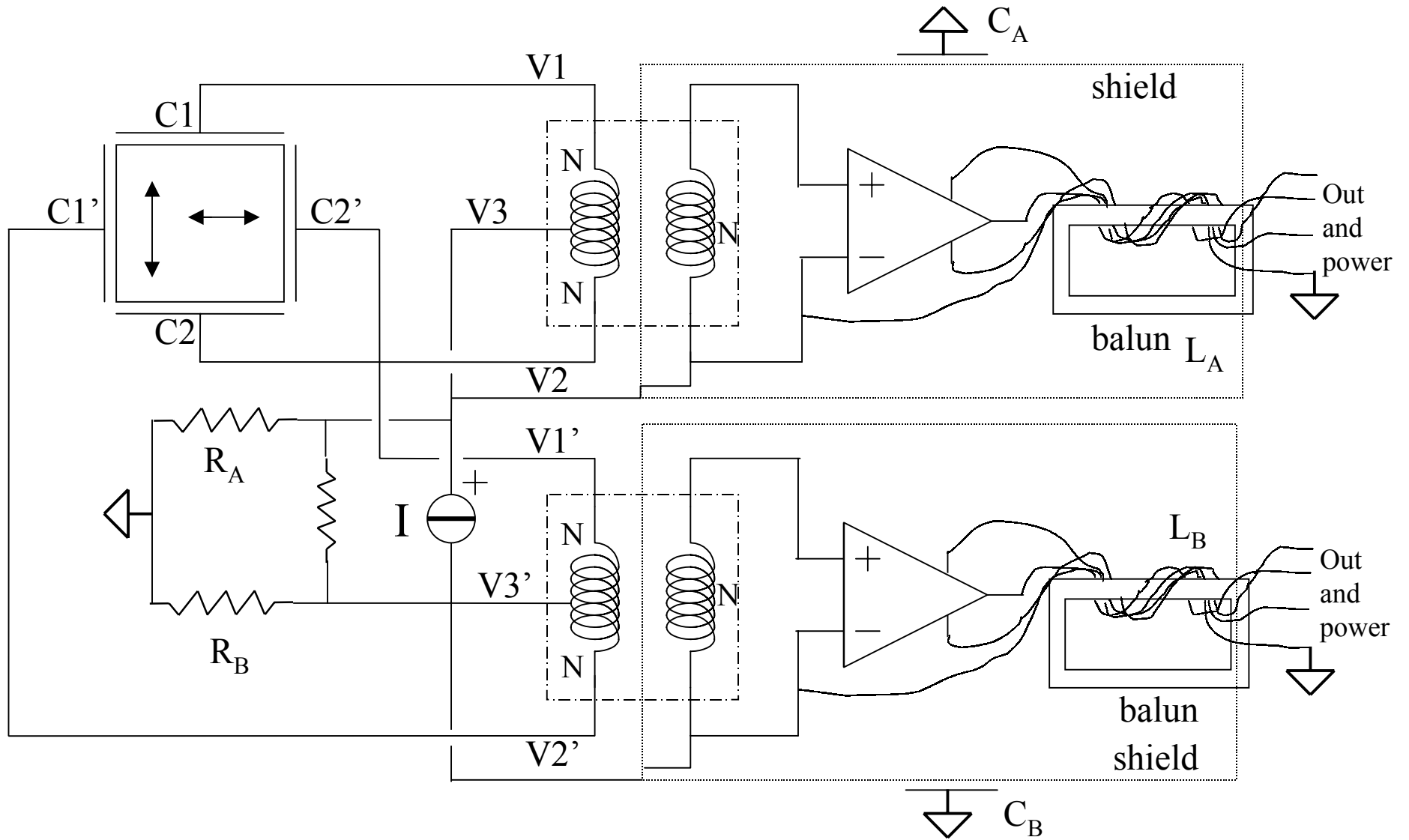

# A floating preamplifier, equivalent circuit for potentials definition.

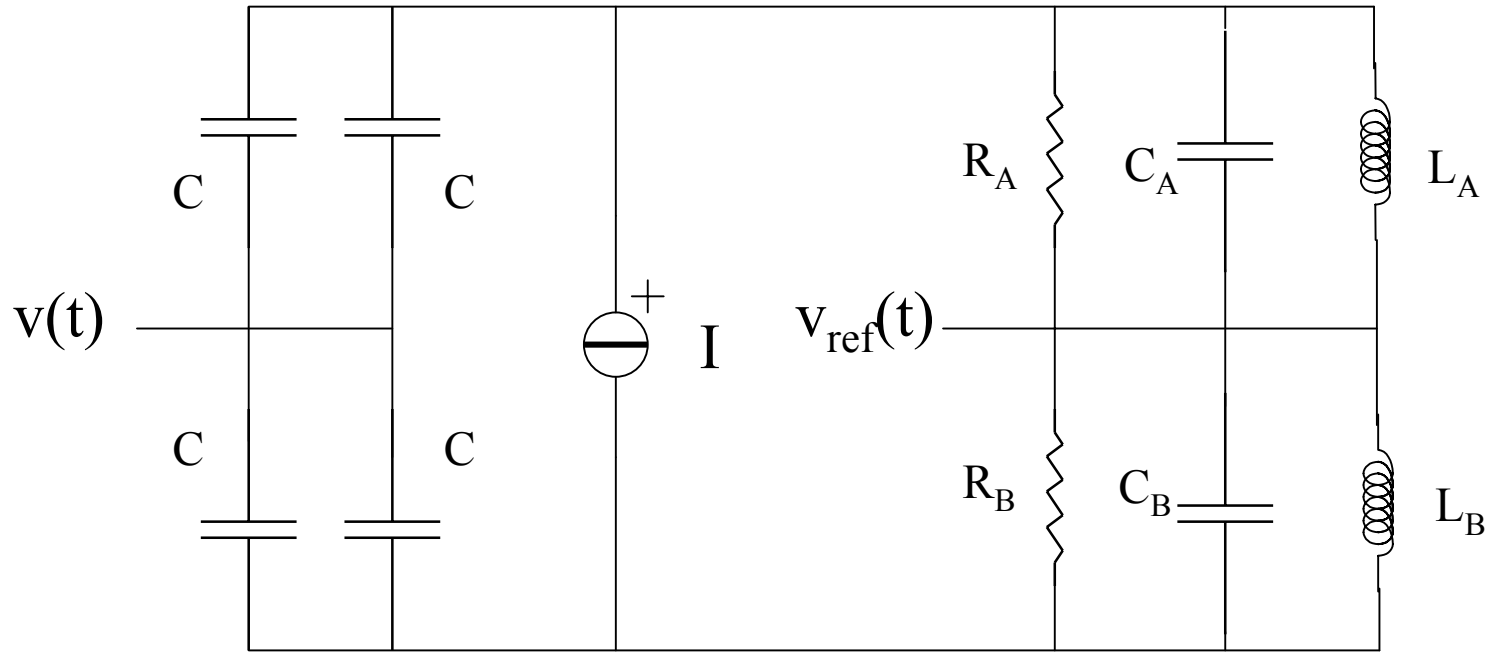

Various options possible for the right RLC circuits: for instance, overdamped (or critically damped) resonant at the bridge frequency or overdamped resonant at a much lower frequency.

CONCLUSION: $C_A$ and $C_B$ could divert a part of generator current, but virtual ground at the proof mass is still possible with the floating "baluned" preamplifier.

Other possible topologies for multidimensional bridges.

# A multifrequency approach.

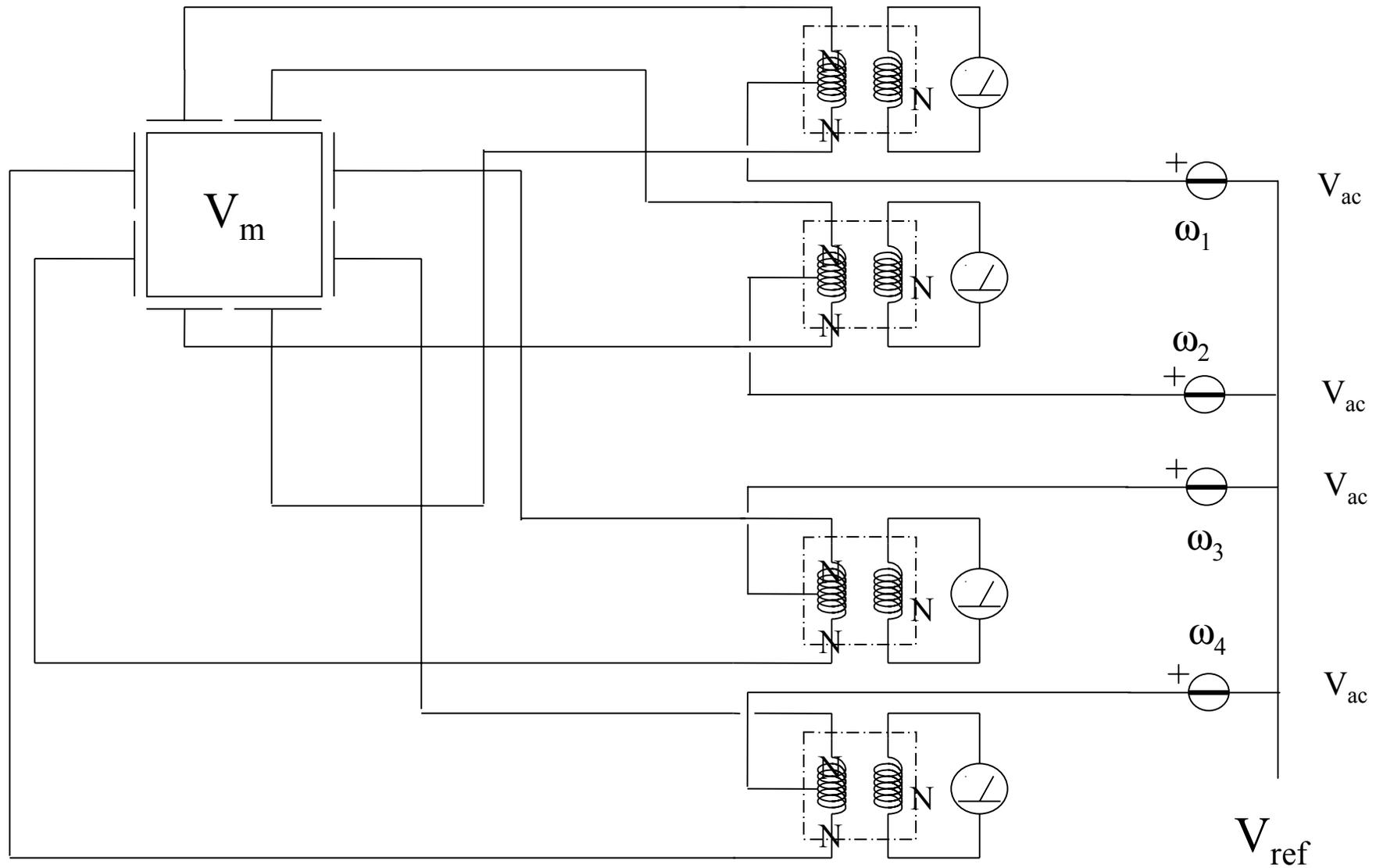

Analysis hint: while considering $\omega_1$ the remaining generators must be considered switched off (shorted). This is superposition of effects.

# Advantages of the multifrequency approach.

1) High symmetry.
2) No injection electrode.
3) No additional stiffness due to injection electrodes.
4) High decoupling among different degrees of freedom because of the multifrequency approach.
5) Maximum use of proof mass area for measurement.
6) Equidistribution of forces due to measurement currents at the equilibrium.
7) Any number of bridges is allowed.

# Disadvantages of the multifrequency approach.

1) High common mode on readout transformers (generator voltage).
2) Readout electrode stiffness higher than conventional configuration because of return currents of other bridges.
3) Proof mass not at zero (reference) potential ($\forall$ t) at the equilibrium, the proof mass averages to zero potential.
4) Necessity of the use of a different frequency for each bridge.
5) Voltage sources required: gain sensitivity to the capacitances of the remaining bridges.

# A single frequency group balanced approach.

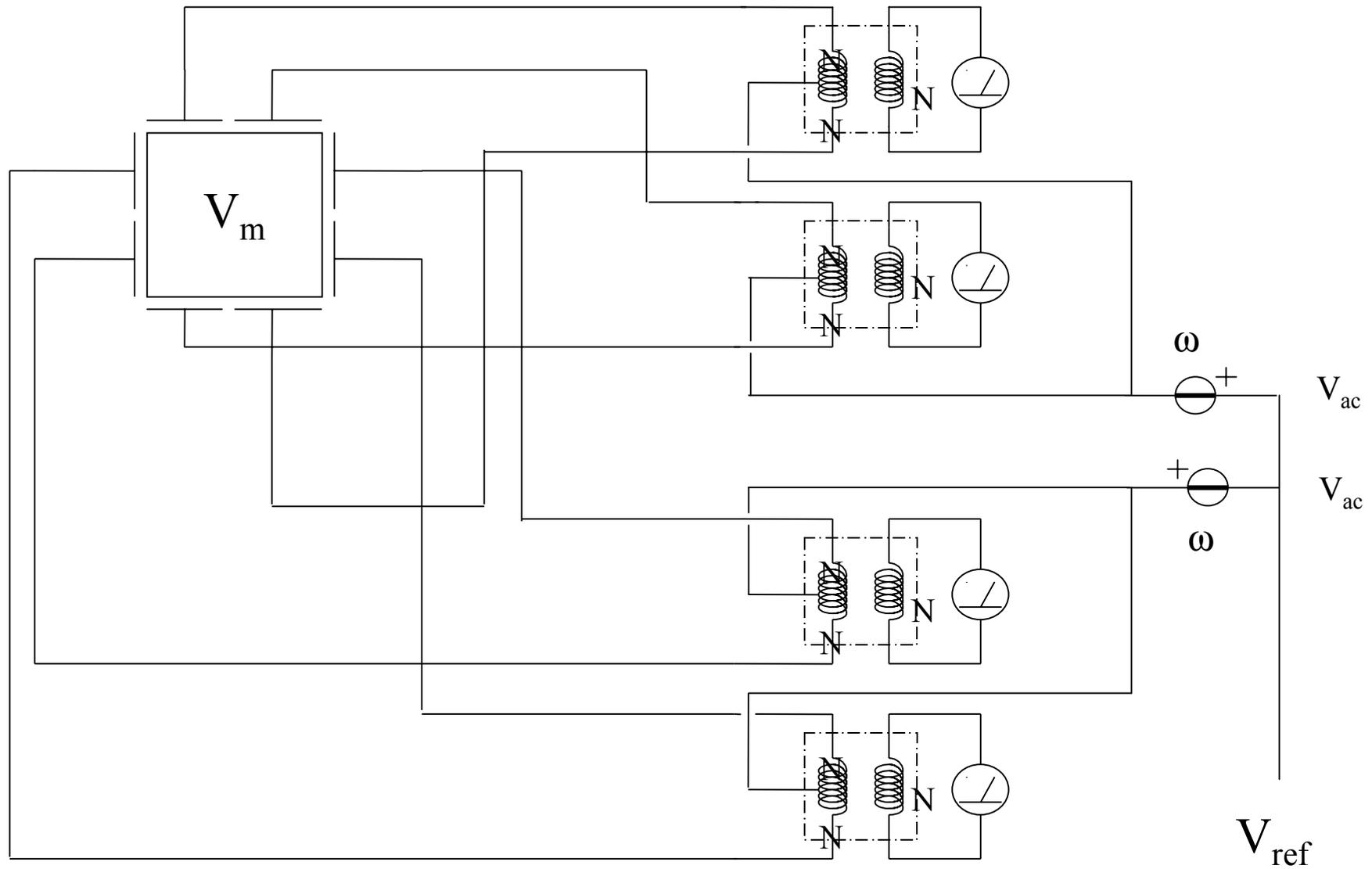

# Advantages of the single frequency group balanced configuration.

1) High symmetry.
2) No injection electrode.
3) No additional stiffness due to injection electrodes.
4) Readout electrode stiffness identical to conventional configuration.
5) Proof mass at zero (reference) potential ($\forall\ t$) at the equilibrium.
6) Proof mass is a *virtual* ground.
7) Maximum use of proof mass area for measurement.
8) Equidistribution of forces due to measurement currents at the equilibrium.
9) Use of a single frequency for multidimensional sensors.

# Disadvantages of the single frequency group balanced configuration.

1) High common mode on readout transformers (generator voltage).
2) Restriction on the number of bridges: only 2N bridges allowed.
3) Voltage sources required: gain sensitivity to the capacitances of the
   · remaining bridges.

# Split electrode DC/AC electrostatic actuator.

*Giorgio Fontana*

*University of Trento*

*dep. Of Materials Engineering*

*september. 2001*

# Principle of Localization of charge.

Example with two degrees of freedom:

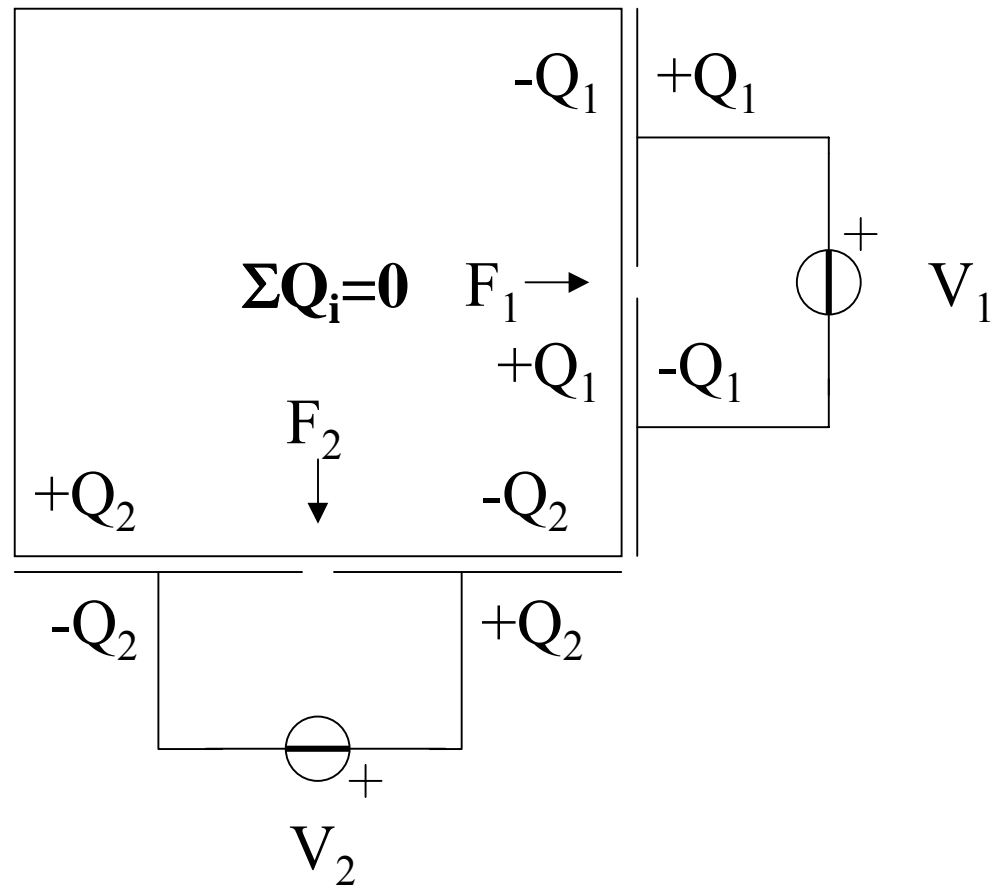

# Splitting of voltage generators and definition of global DC potential.

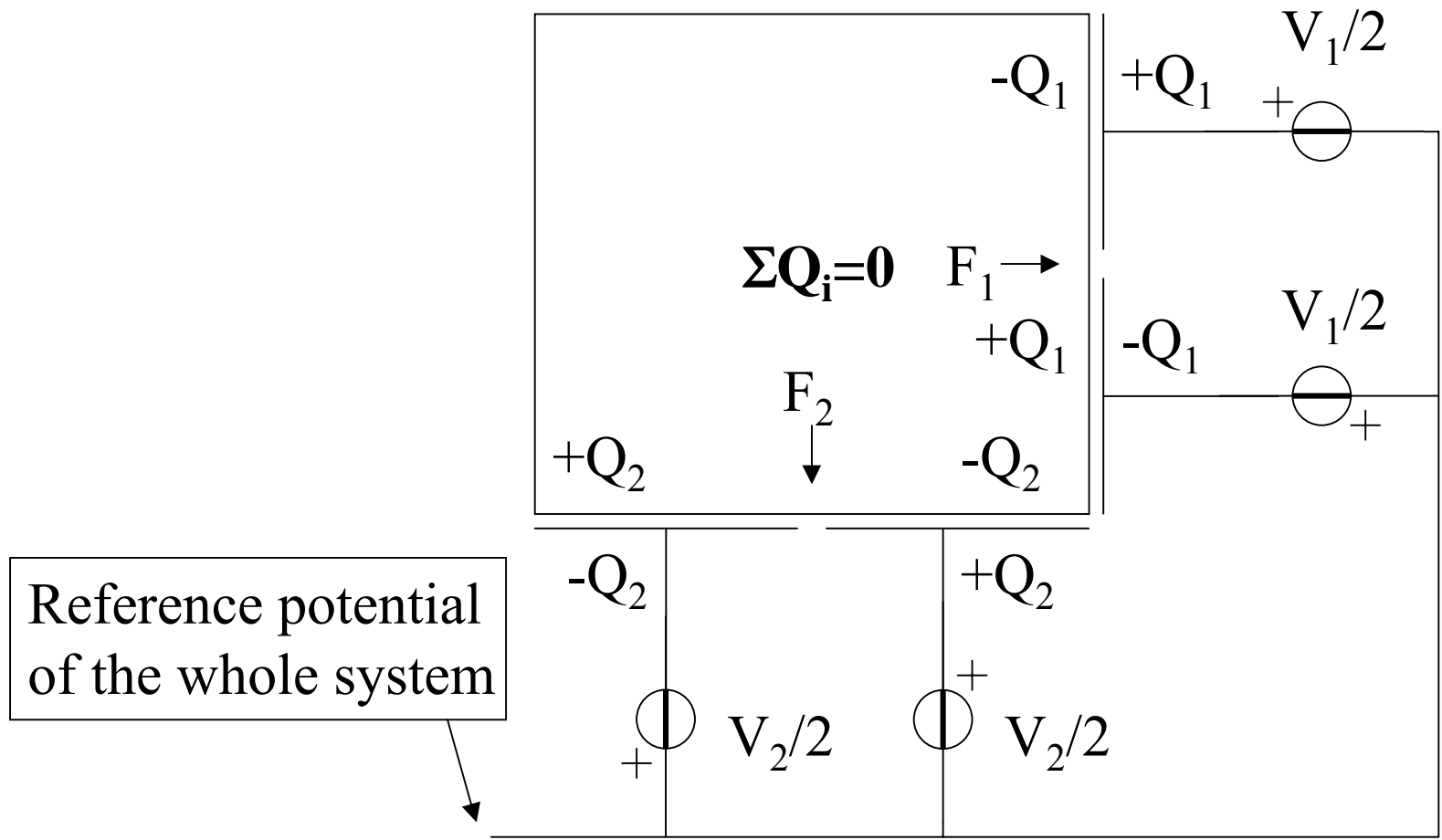

# Advantages of the proposed configuration.

1) High symmetry.
2) No injection electrode.
3) No additional stiffness due to injection electrodes.
4) Stiffness localized in the active electrodes
   for instance: no force on $x$ = no stiffness on $x$.
5) Proof mass at zero (reference) potential ($\forall$ t) at the equilibrium.
6) Proof mass is a *virtual* ground.
7) Maximum use of proof mass area for actuation.
8) Independent control of each degree of freedom.
9) AC or DC actuation is possible.
10) Possible use of a single frequency for AC multidimensional actuators.

# Disadvantages of the proposed configuration.

1) Four electrodes for each degree of freedom for push-pull operations, compared to two electrodes for more conventional configurations which might employ an injection electrode.

# Split electrode AC electrostatic actuator with zero stiffness.

*Giorgio Fontana*

*University of Trento*

*dep. Of Materials Engineering*

*November. 2001*

# Force and stiffness on electrodes of a plane capacitor with charge as a controlled variable.

Capacitance of a plane capacitor: $$C = \frac{\varepsilon S}{x}$$

S is the electrode area, x the distance between the electrodes and $\varepsilon$ is the permittivity.
Using some well known relations we have:

$$Q = CV, \quad E = \frac{1}{2} C V^2 = \frac{1}{2} Q^2 \frac{x}{\varepsilon S}$$

$$F = \frac{\partial E}{\partial x} = \frac{1}{2} \frac{Q^2}{\varepsilon S}$$

If Q does not depend on, x then the stiffness: $k = \dfrac{\partial F}{\partial x} = 0$

Charge on the electrodes of a plane capacitor in an AC current controlled regime (with frequency $f=\omega/2\pi$).

$$Q = CV \qquad C = \frac{\varepsilon S}{x} \quad \text{(we do not use this, we simply observe that only C depend on x)}$$

$$I = \omega CV \qquad \omega C \text{ is the modulus of capacitor admittance.}$$

$$Q = \frac{I}{\omega} \qquad F = \frac{\partial E}{\partial x} = \frac{I^2}{2\varepsilon S \omega^2}$$

For a controlled current driving of the actuator, the Charge and the Force do not depend on the distance between the electrodes. Moreover the stiffness is always zero.

Here we ignore the phase relation between current and voltage because it has no effect on the force. The Force can be computed with the RMS of I.

# Example with two degrees of freedom:

The four equal resistors must have a resistance much higher than the reactance of the capacitors of the actuator. Their role is the definition of the DC potential of the proof mass and can be shared with the measurement bridges.

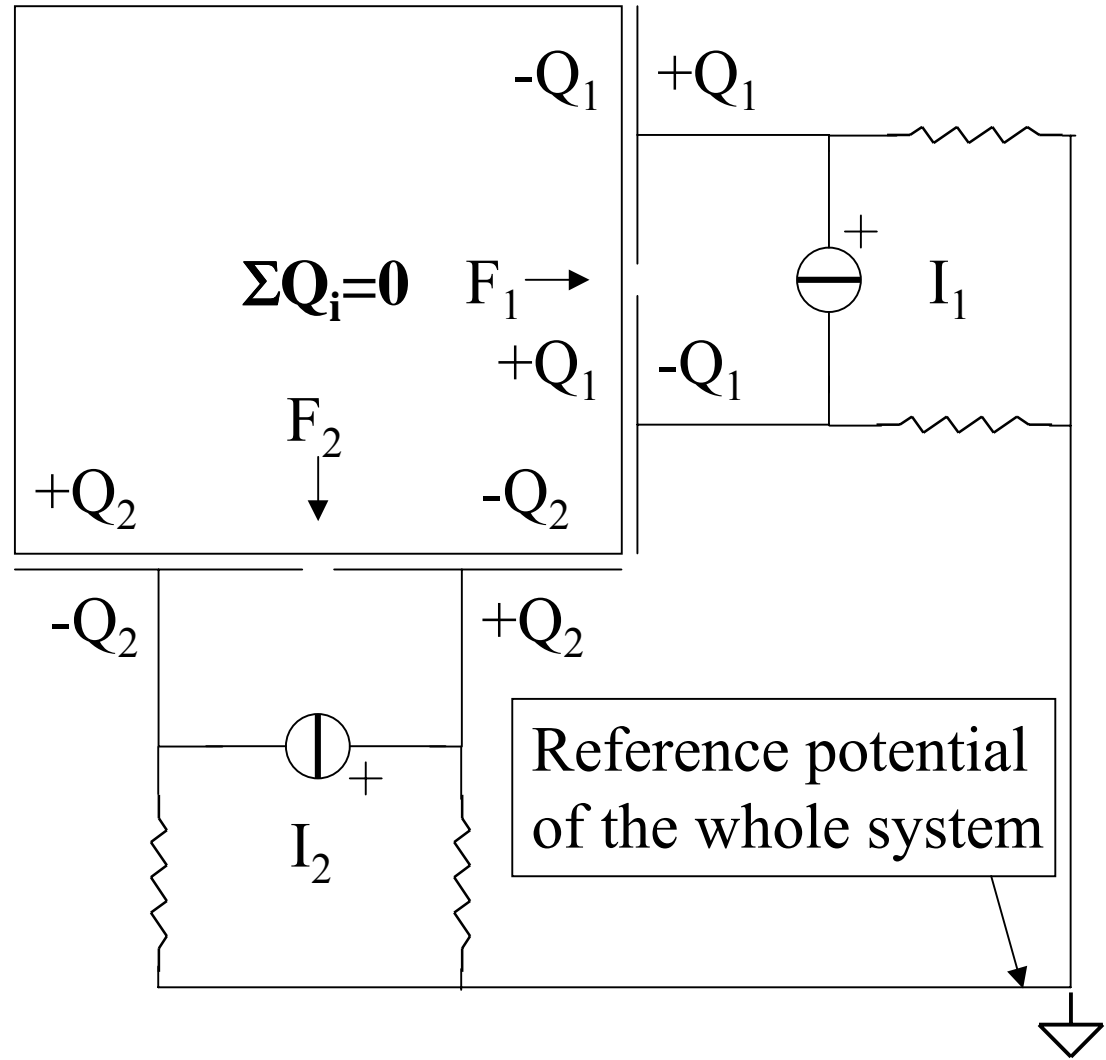

# Major technical issues of current drive:

1) Ability of producing accurate sine currents with small amplitude.

2) Ability to send the generated current to the actuator electrodes without dispersion.

Point 2 needs the development of techniques capable of compensating or shielding parasitic capacitance.
Examples follow.

# Compensating parasitic capacitances (step 0): (a possible technique)

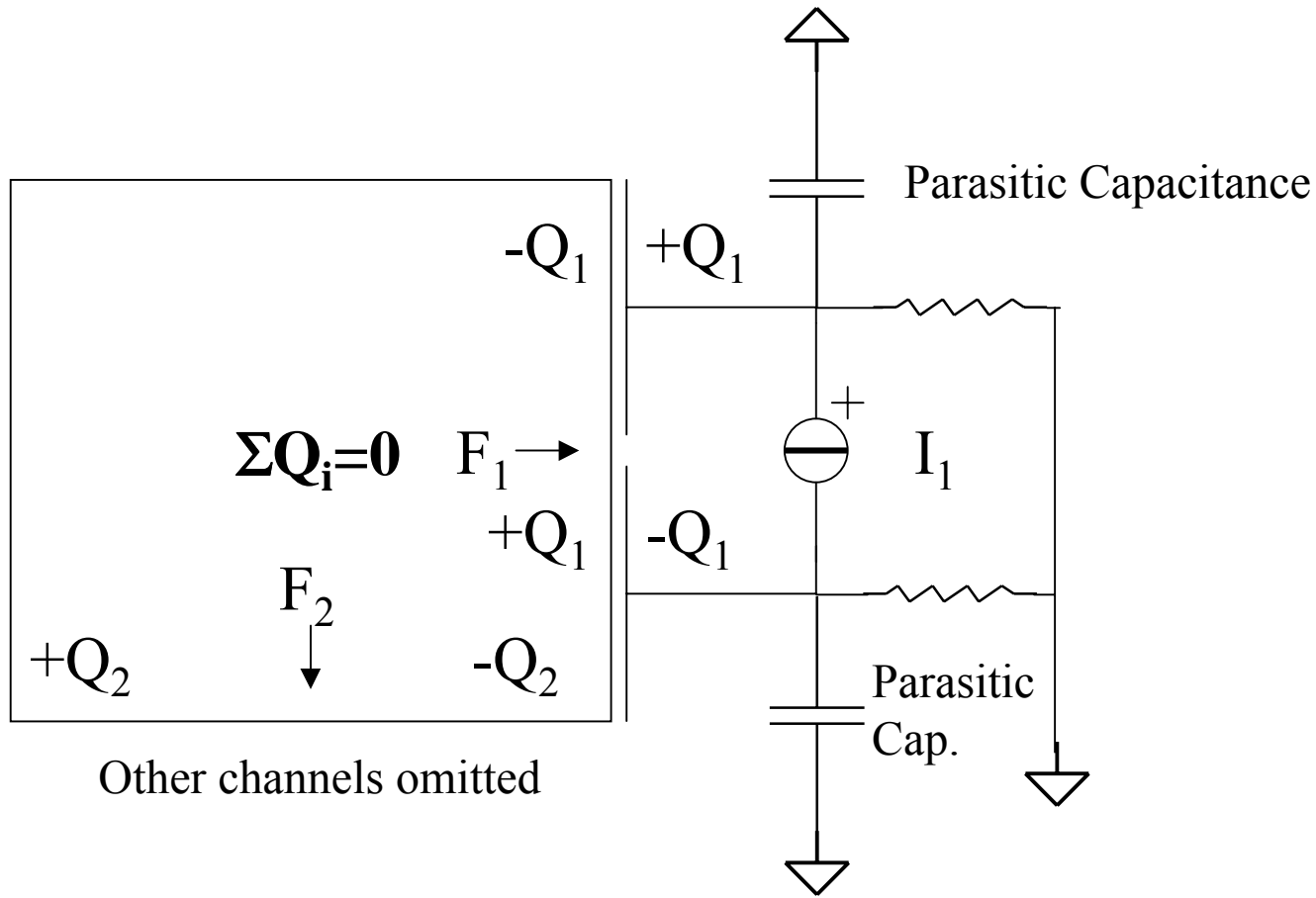

$\Sigma Q_i = 0$

Other channels omitted

# Compensating parasitic capacitances (step 1):

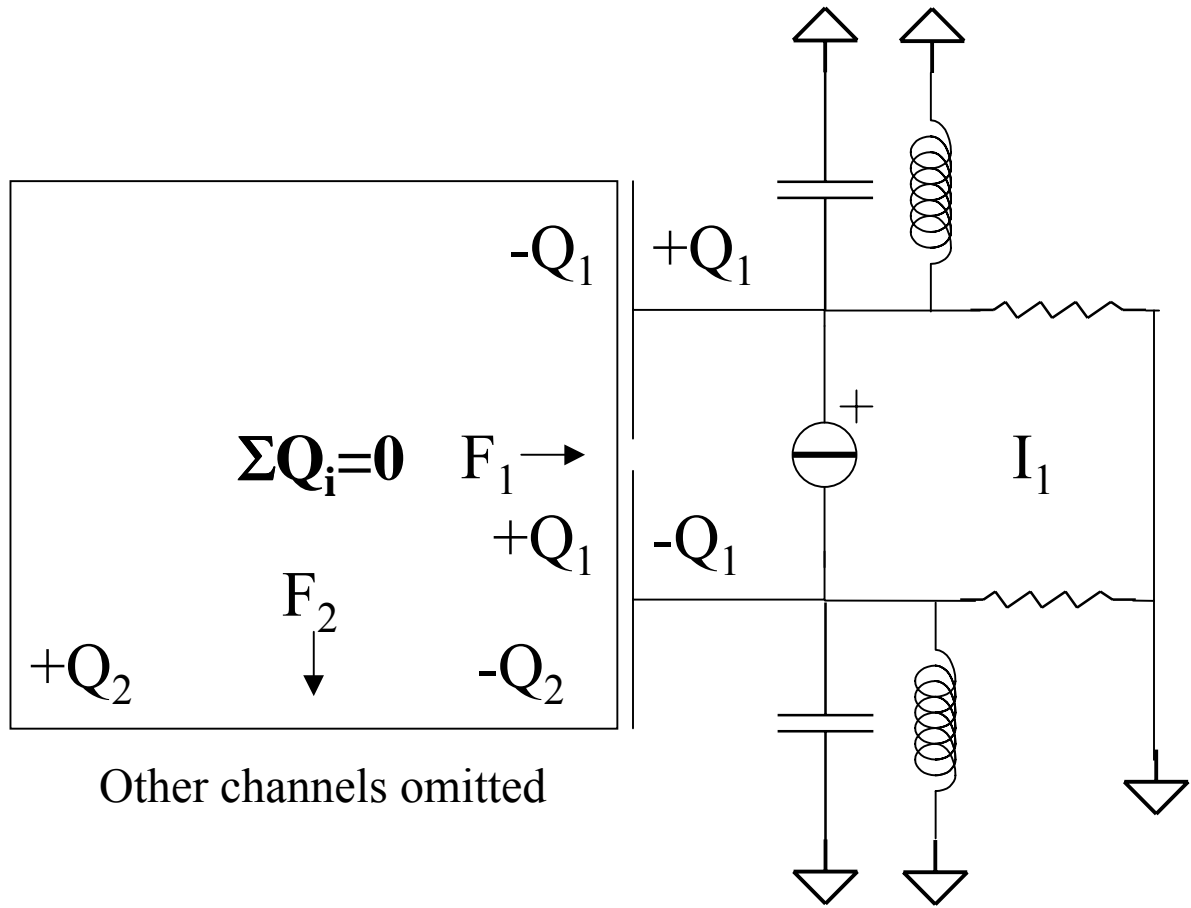

Other channels omitted

Resonating parasitic capacitances at the excitation frequency.

# Compensating parasitic capacitances (step 2):

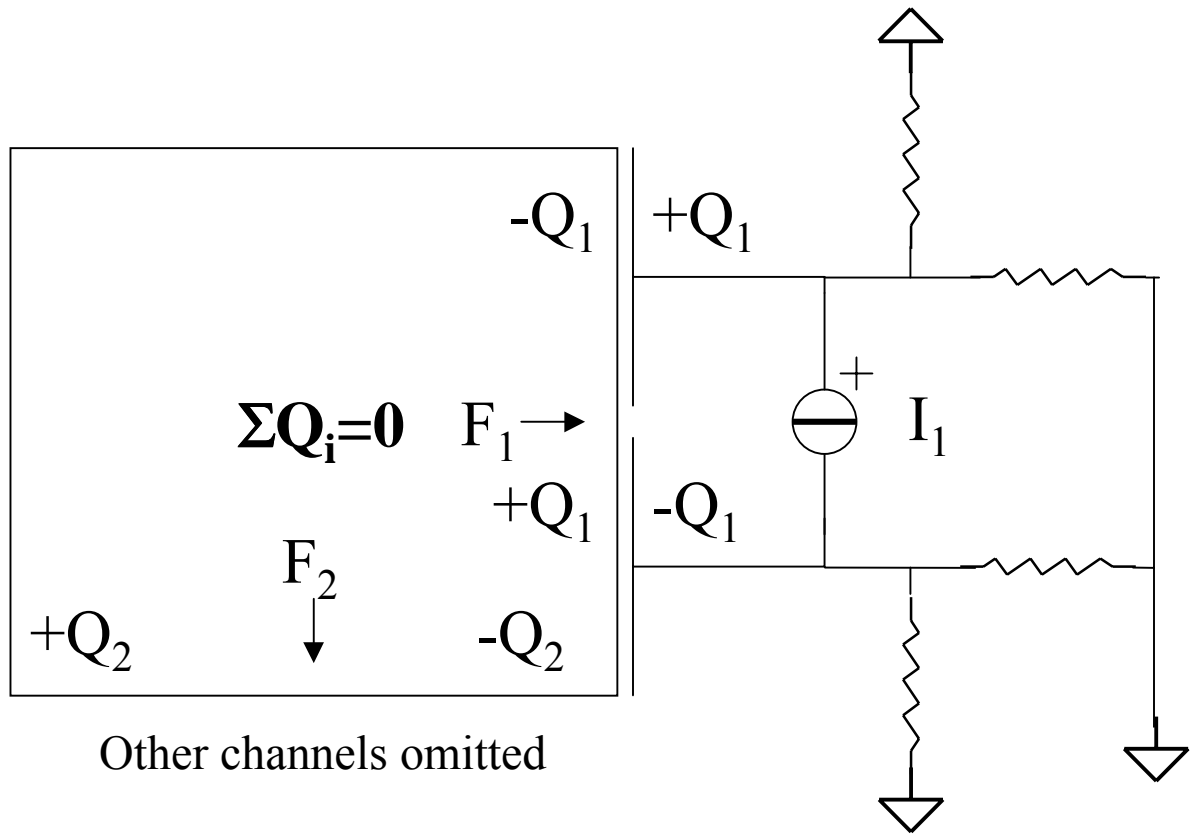

Other channels omitted

Representing LC losses at the resonance.

# Compensating parasitic capacitances (step 3):

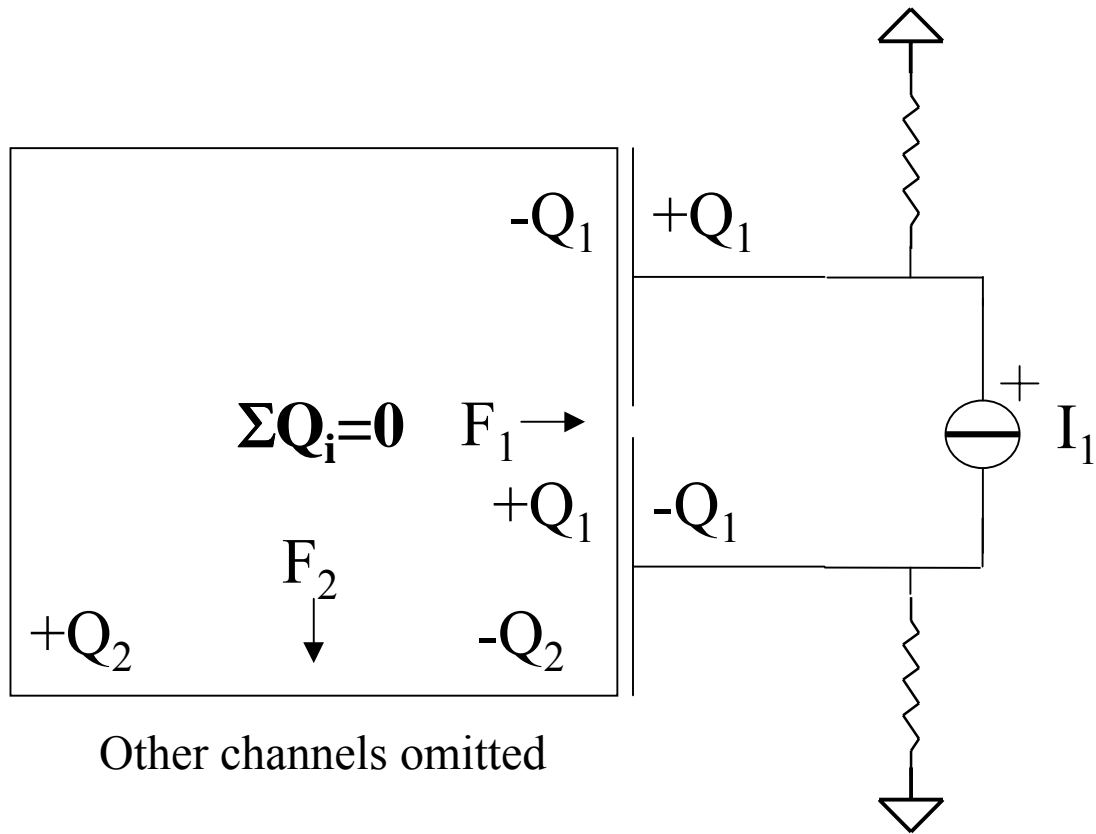

Equivalent circuit.

# Compensating parasitic capacitances (step 4):

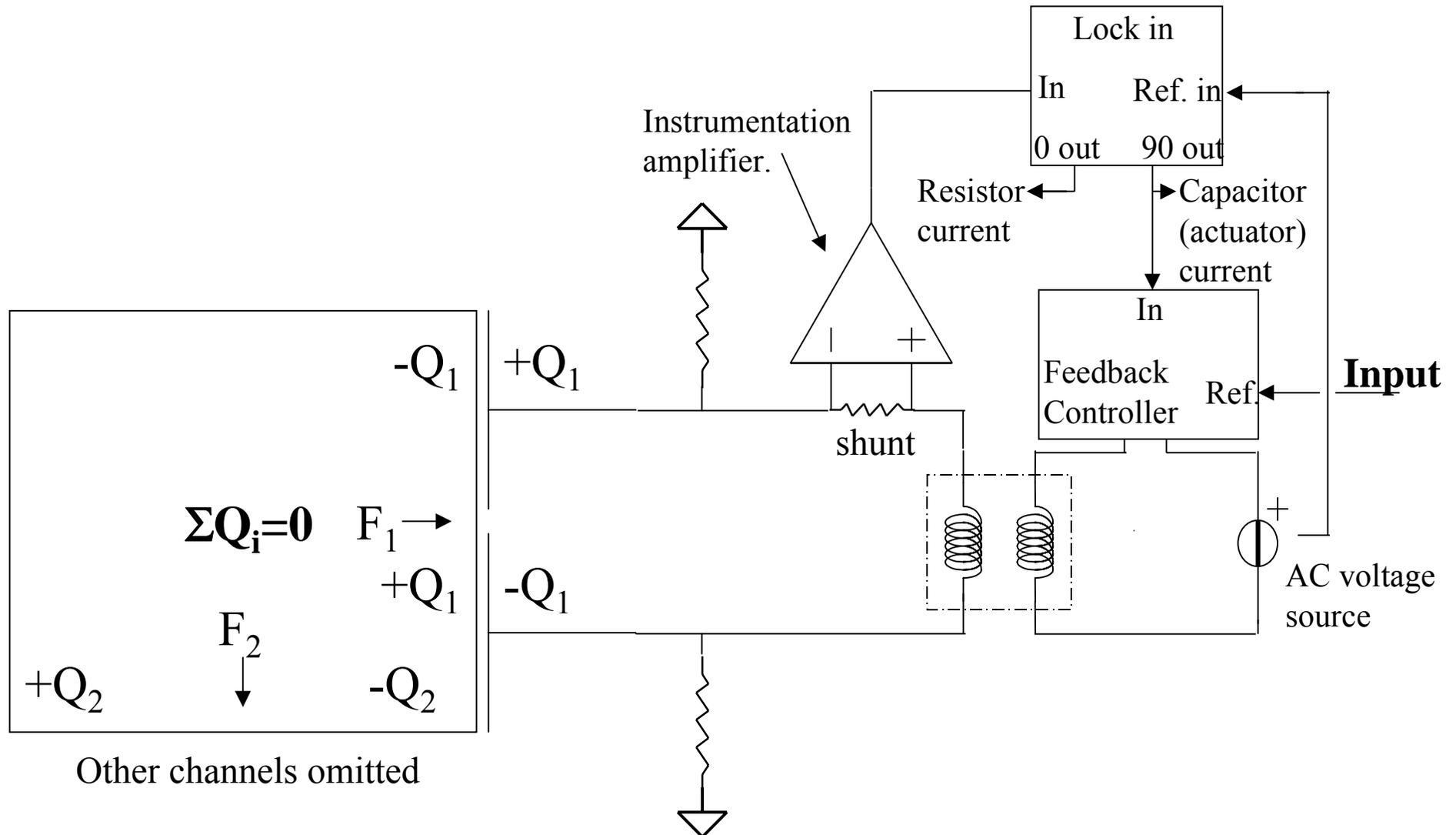

Producing the current by a Phase Sensitive Detector controlled voltage source with a shunt.

# Bootstrapping parasitic capacitances with voltage controlled guard electrodes.
## (with $C_{in}$ of OpAmps $<<$ C of Actuator):

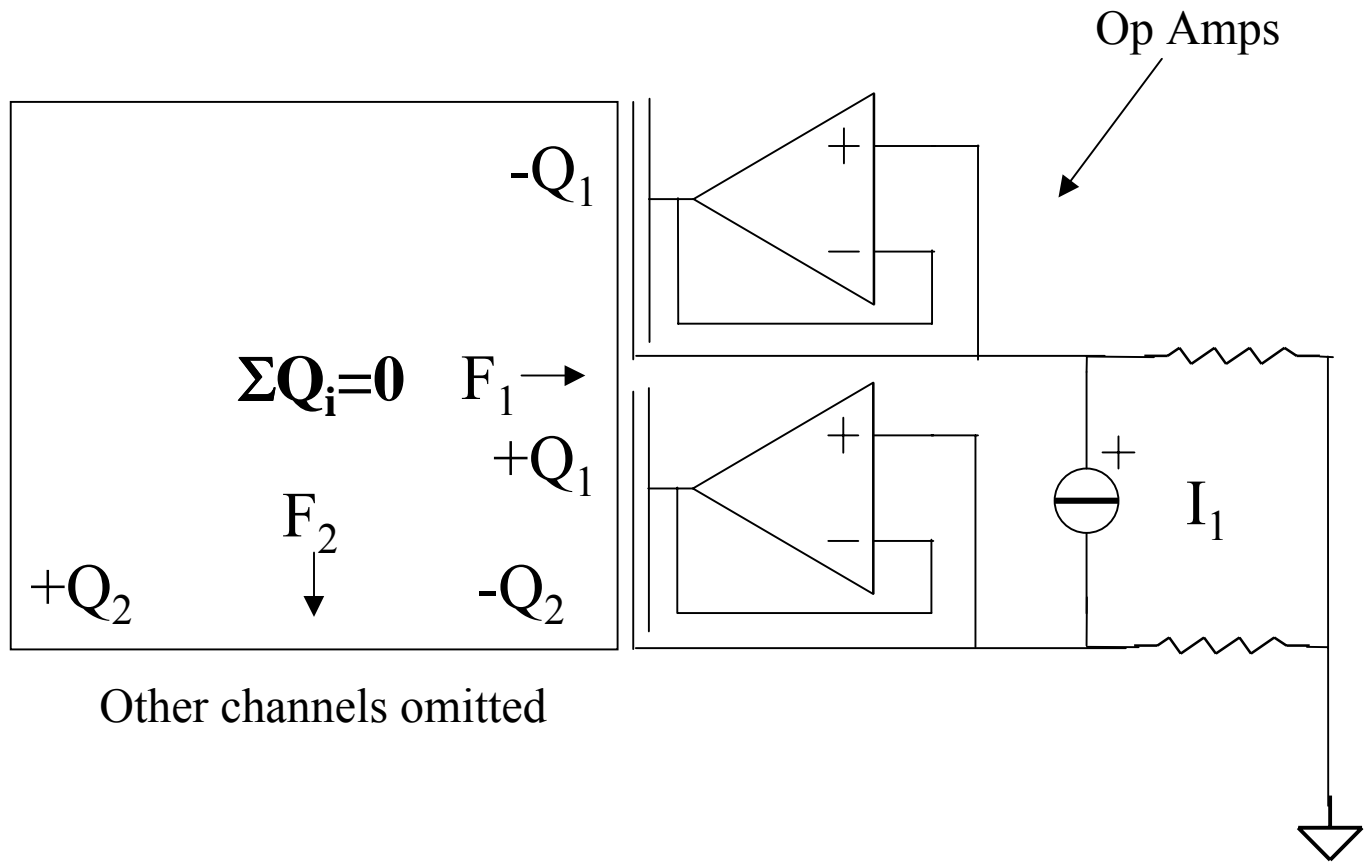

Other channels omitted

# Feedback controlled compensated current source. (the parasitic current is subtracted from measurement):

- Parasitic Capacitance
- Instrumentation amplifiers, not OP-AMPs.
- Rectifier
- In Feedback Controller — Ref. ← **Input**
- $-Q_1$ | $+Q_1$
- shunt
- $\Sigma Q_i = 0$  $F_1 \rightarrow$
- Compensation Voltage Divider
- $+Q_1$ | $-Q_1$
- $F_2 \downarrow$
- $+Q_2$  $-Q_2$
- AC voltage source
- Parasitic Capacitance.
- Other channels omitted
- Consider symmetric parasitic capacitances

# Broadband AC Compensated Current Injection (CCI) circuit.

**Input** V1

OP-AMP

$I1 = V1\omega(C1+C2)$

$I2 = V2\omega(C3+C4)$

G=1

Parasitic Capacitance

C1  C2  C1

$-Q_1$ | $+Q_1$

I1  C2

$\Sigma Q_i = 0$  $F_1 \rightarrow$

$+Q_1$ | $-Q_1$

$F_2$

$+Q_2$  $-Q_2$

I2

The same as above with C3 and C4 in place of C1 and C2.

V2 **Input**

C3

Parasitic Capacitance

Other channels omitted

© Giorgio Fontana

# Advantages of the split electrode AC electrostatic current driven actuator.

1) High symmetry.
2) No injection electrode.
3) No additional stiffness due to injection electrodes.
4) Zero stiffness on $x_i$.
5) Force does not depend on $x_i$.
6) Proof mass at zero (reference) potential ($\forall\ t$) at the equilibrium.
7) Proof mass is a *virtual* ground.
8) Maximum use of proof mass area for actuation.
9) Independent control of each degree of freedom.
10) Possible use of a single frequency for AC multidimensional actuators.

# Disadvantages of the split electrode AC electrostatic current driven actuator.

1) Four electrodes for each degree of freedom for push-pull operations, compared to two electrodes for more conventional configurations which might employ an injection electrode.

# Combination of a double bridge and a split electrode actuator.

The double bridge in its different possible configurations can be combined with the split electrode actuator, voltage or current controlled, for a complete system, without loosing the advantages of each configuration. In the following example the bridge frequency is much higher than actuator frequency (if an AC actuator schema is employed).

# The bidimensional bridge and a single voltage driven split electrode actuator.

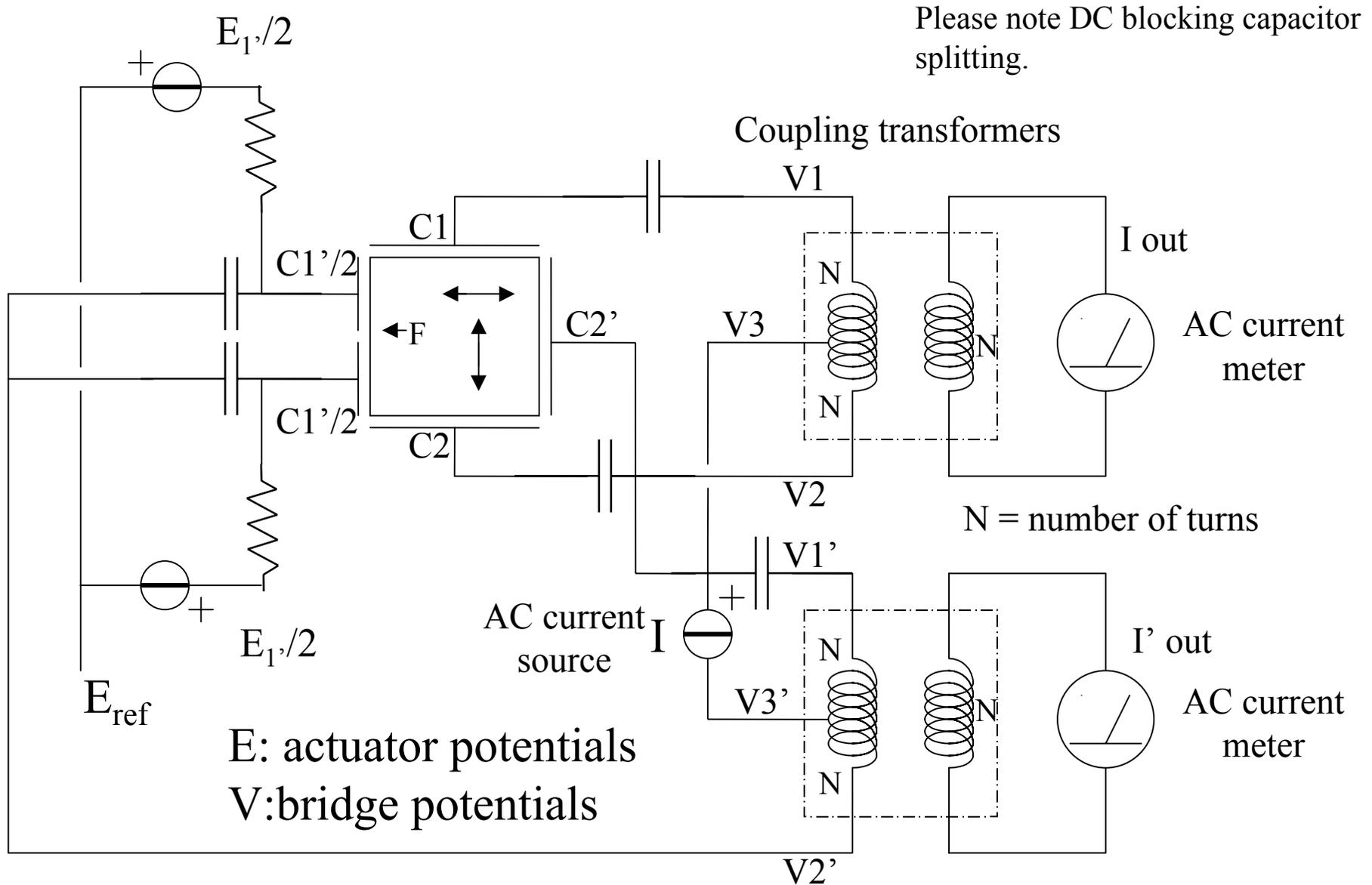